\documentclass[lettersize,journal]{IEEEtran}
\usepackage{amsmath,amsfonts}
\usepackage{algorithmic}
\usepackage{algorithm}
\usepackage{array}
\usepackage{textcomp}
\usepackage{url}
\usepackage{verbatim}
\usepackage{graphicx}
\usepackage{cite}
\usepackage{float}
\usepackage{multirow}
\usepackage{booktabs}
\usepackage{xcolor}
\usepackage{graphicx}
\usepackage[tight,footnotesize]{subfigure}
\begin{document}


\title{GLST: Defending Confidence-Driven V2X Collaborative Perception Against Stealthy Multi-Attacker Feature Injection}

\author{Ji He,~\IEEEmembership{Member,~IEEE,}
       Ying Wang,
       Lijie Zheng,~\IEEEmembership{Graduate Student Member,~IEEE,} 
       Xinghui Zhu,~\IEEEmembership{Member,~IEEE,}
      Yulong Shen,~\IEEEmembership{Member,~IEEE,}
      Xiaohong Jiang,~\IEEEmembership{Senior~Member,~IEEE}
\thanks{J. He, Y. Wang, L. Zheng, X, Zhu, Y. Shen and X. Jiang are with the School of Computer Science and Technology, Xidian University, Xi'an, China (jihe@xidian.edu.cn; 18223927854@163.com; lijzheng@stu.xidian.edu.cn; xhzhu@xidian.edu.cn; ylshen@mail.xidian.edu.cn; jiang@fun.ac.jp).}
}

\maketitle

\begin{abstract}
Collaborative perception (CP) improves autonomous driving perception by allowing connected vehicles to exchange intermediate features through V2X communication. Among existing approaches, confidence-driven sparse communication has become an important design paradigm because it reduces bandwidth consumption by transmitting only perception-critical spatial regions. However, this design also introduces a new security risk: once a collaborator is compromised, malicious features located in high-confidence or ego-uncertain regions may be preferentially selected and amplified during feature fusion. In this paper, we investigate this vulnerability using Where2comm as a representative confidence-driven collaborative perception framework. We first verify that the proposed Pretend Benign attack can effectively exploit the spatial confidence mechanism of Where2comm by injecting stealthy perturbations into uncertain yet perception-critical regions, significantly degrading 3D object detection while maintaining benign-like feature characteristics.

Beyond this specific attack-framework pair, we further identify a more general weakness of existing trust-based defenses: many of them rely primarily on a single consistency signal, making them vulnerable to multi-attacker scenarios where malicious agents can form a pseudo-consensus and bias trust estimation. To address this problem, we propose Global-Local Structural Trust (GLST), a lightweight defense framework that evaluates collaborator reliability from three complementary perspectives: global feature consistency, multi-scale local residual consistency, and structural consistency with ego-side semantic topology. The resulting trust scores are incorporated into the feature fusion process to suppress unreliable collaborators. Experiments on the OPV2V dataset demonstrate that GLST achieves competitive performance under single-attacker Pretend Benign attacks and provides substantially stronger robustness under multi-attacker settings. In the four-attacker Pretend Benign scenario, GLST maintains 0.69 AP@0.3, whereas existing single-signal defenses suffer severe degradation. GLST also remains effective against gradient-based attacks such as PGD, suggesting that multi-level trust modeling is essential for securing confidence-driven collaborative perception.
\end{abstract}



\section{Introduction}

\IEEEPARstart{C}{ollaborative} perception (CP) has emerged as an important technique for improving autonomous driving perception under occlusion, limited sensing range, and incomplete local observations~\cite{graham20183d,xia2023coin}.
By leveraging Vehicle-to-Everything (V2X) communication, multiple connected autonomous vehicles (CAVs) can exchange perception information and aggregate complementary observations into a more complete environmental representation~\cite{chen2019cooper,chen2019f}.
Among different CP paradigms, feature-level collaboration has attracted increasing attention because it preserves rich spatial information while avoiding the prohibitive communication cost of raw-data sharing ~\cite{ye2020cooperative,yang2024density}.

Recent CP systems have further improved communication efficiency through selective and confidence-driven communication~\cite{yoo2025learning,wang2023core,yang2023how2comm,yang2023spatio}. Where2comm~\cite{hu2022where2comm}, as a representative confidence-driven sparse communication framework, transmits only perception-critical feature regions according to spatial confidence maps. This design substantially reduces bandwidth consumption while maintaining competitive 3D object detection performance.
However, the same mechanism may also introduce a new security risk.
Since perception-critical regions are preferentially selected for communication and then involved in feature fusion, malicious features injected into these regions may be propagated and amplified by the CP pipeline.
Therefore, confidence-driven sparse communication, while efficient, can unintentionally enlarge the attack surface of feature-level CP systems.

The open communication architecture of CP makes this threat particularly concerning.
Once a collaborator is compromised, it may transmit carefully crafted adversarial features to the ego vehicle.
If the ego vehicle fails to identify unreliable collaborators or suppress their injected features, the final perception output may be manipulated, leading to missed detections, false positives, or shifted bounding boxes.
Recent stealthy feature-injection attacks, such as the Pretend Benign (PB) attack~\cite{lin2025pretend}, further intensify this challenge.
PB optimizes adversarial features to resemble benign feature distributions while still misleading downstream detection heads, making malicious features difficult to detect through conventional anomaly filtering or consistency checking mechanisms.

Existing CP defenses, including ROBOSAC~\cite{li2023among}, LUCIA~\cite{wang2025threat}, and MADE~\cite{zhao2024made}, improve robustness by estimating collaborator reliability and suppressing suspicious information.
However, many of these methods rely mainly on a single global or pairwise consistency signal.
Such a design can be fragile under stealthy and multi-attacker scenarios.
When multiple malicious agents jointly participate in CP, they may form a pseudo-consensus in the feature space, which increases their mutual consistency and biases trust estimation.
As a result, conventional consistency-based defenses may assign overly high reliability to coordinated attackers and fail to prevent corrupted feature fusion.

In this paper, we focus on the security of confidence-driven sparse collaborative perception.
We use Where2comm as a representative framework because it combines spatial confidence-based communication with attention-based feature fusion.
Our goal is not to design a defense tailored only to Where2comm, but to study a broader security issue in confidence-driven feature-level CP: malicious features located in perception-critical regions may be preferentially transmitted and amplified during fusion.
We systematically analyze this vulnerability under PB attacks and show that stealthy feature injection can significantly degrade cooperative 3D object detection while maintaining benign-like feature characteristics. To address this problem, we propose Global-Local Structural Trust (GLST), a lightweight defense framework for secure confidence-driven collaborative perception. Instead of relying on a single consistency signal, GLST jointly evaluates collaborator reliability from three complementary perspectives: global feature consistency, multi-scale local residual consistency, and structural consistency with ego-side semantic topology.
The resulting trust scores are used to reweight the feature fusion process, thereby suppressing low-trust collaborators at the feature level without requiring complex reconstruction, iterative sampling, or additional communication procedures. Our main contributions are summarized as follows:
\begin{itemize}
\item We identify and empirically analyze a critical security vulnerability in confidence-driven sparse CP. Using Where2comm as a representative framework, we show that confidence-driven communication and attention-based fusion, while effective in reducing communication overhead, may unintentionally amplify adversarial feature propagation. In particular, stealthy feature-injection attacks such as PB can exploit perception-critical spatial regions by concentrating perturbations on uncertain or locally weakly perceived areas, thereby degrading cooperative 3D object detection while preserving benign-like feature characteristics.

\item We reveal that existing consistency-based trust defenses are fragile under multi-attacker CP scenarios. Since many defenses estimate collaborator reliability mainly from a single global or pairwise consistency signal, multiple malicious agents can reinforce each other and form a pseudo-consensus in the feature space. This biased consensus weakens the discriminability between benign and malicious collaborators and may cause conventional trust estimation mechanisms to assign overly high reliability to coordinated attackers. 

\item We propose GLST, a lightweight defense framework for secure confidence-driven collaborative perception. GLST jointly evaluates collaborator reliability from three complementary perspectives: global feature consistency, multi-scale local residual consistency, and structural consistency with ego-side semantic topology. The resulting trust scores are used to reweight the feature fusion process, thereby suppressing low-trust collaborators and mitigating adversarial interference at the feature level.

 \item We conduct extensive experiments on the OPV2V dataset ~\cite{xu2022opv2v} under diverse adversarial settings, including single-attacker and multi-attacker PB attacks, as well as gradient-based attacks such as PGD~\cite{madry2017towards}. The results show that GLST achieves performance comparable to representative defenses in single-attacker scenarios while providing stronger robustness under multi-attacker settings, demonstrating the importance of multi-level trust modeling for secure CP.      
\end{itemize}


\section{Related Work}
\label{sec:related_work}


\subsection{Adversarial Attacks Against CP}

Adversarial attacks against perception systems were first studied in single-agent vision scenarios.
Representative methods, such as pixel-level perturbations~\cite{xie2017adversarial} and Projected Gradient Descent (PGD)~\cite{madry2017towards}, demonstrate that carefully optimized perturbations can mislead deep perception models. Subsequent studies further extended adversarial attacks to physical-world perception~\cite{hu2021naturalistic,li2018exploring}, showing that printable perturbations, adversarial patches, and natural camouflage patterns can threaten real-world perception systems.

Compared with single-vehicle perception, CP introduces an additional communication-layer attack surface because the ego vehicle relies on information shared by external collaborators ~\cite{liu2020when2com,xu2022v2x,yang2023spatio}. A compromised agent can therefore manipulate the collaborative fusion process by perturbing shared features, falsifying pose information, or fabricating object-level messages. Tu \textit{et al.}~\cite{tu2021adversarial} showed that controlling only a few malicious agents can cause severe missed detections and false positives by attacking shared feature representations. AdvGPS~\cite{li2024advgps} disrupts feature alignment by attacking pose information, while Zhang \textit{et al.}~\cite{zhang2024data} model communication-layer attacks as object spoofing and object removal. More recently, stealthy feature-injection attacks have posed a stronger challenge to CP security. The PB attack~\cite{lin2025pretend} optimizes adversarial features to preserve benign-like characteristics while still misleading downstream detection heads. Compared with conventional attacks that introduce more visible abnormal perturbations, PB is harder to detect using simple anomaly filtering or consistency checking. For confidence-driven sparse communication frameworks such as Where2comm, this threat becomes more serious because malicious perturbations can be concentrated on perception-critical regions that are more likely to be transmitted and fused. Therefore, PB provides a representative threat model for analyzing the security of confidence-driven feature-level CP.

\subsection{Defenses Mechanisms for Secure CP}

To improve the robustness of CP under compromised communication environments, existing defenses mainly attempt to identify unreliable collaborators or suppress suspicious information during fusion. ROBOSAC~\cite{li2023among} follows a sampling-based strategy, which searches for a malicious-free collaborator subset by comparing sampled collaborative predictions with the ego vehicle's independent perception results. Another line of work constructs trust scores directly from intermediate features.LUCIA~\cite{wang2025threat} assigns soft trust weights based on pairwise feature distances among agents and suppresses suspicious collaborators during feature fusion. Anomaly-detection-based defenses provide another way to evaluate collaborator reliability. MADE~\cite{zhao2024made} performs a dual-hypothesis test for each collaborator by jointly considering matching loss and collaborative reconstruction loss. Overall, existing CP defenses are largely built upon a single source of evidence, e.g., sampling-based consensus, global feature distance, or anomaly scores. This design may be insufficient against compound threats that combine distribution mimicry with multi-attacker collaboration.


\section{Preliminaries}

\subsection{Confidence-Driven Collaborative Perception in V2X}\label{sec:Collaborative_Perception_Framework}

CP enables multiple CAVs to share complementary observations through V2X communication, thereby mitigating the limited field of view and occlusion problems of single-agent perception~\cite{wu2024commonsense,lang2019pointpillars,yan2018second}. Among existing CP paradigms, intermediate feature fusion has become a widely adopted solution because it retains rich semantic and geometric cues while requiring much less communication bandwidth than raw point-cloud sharing. This trade-off makes feature-level collaboration particularly suitable for real-time 3D object detection in multi-agent driving scenarios.

\begin{figure}[t]
    \centering
    \includegraphics[width=0.45\textwidth]{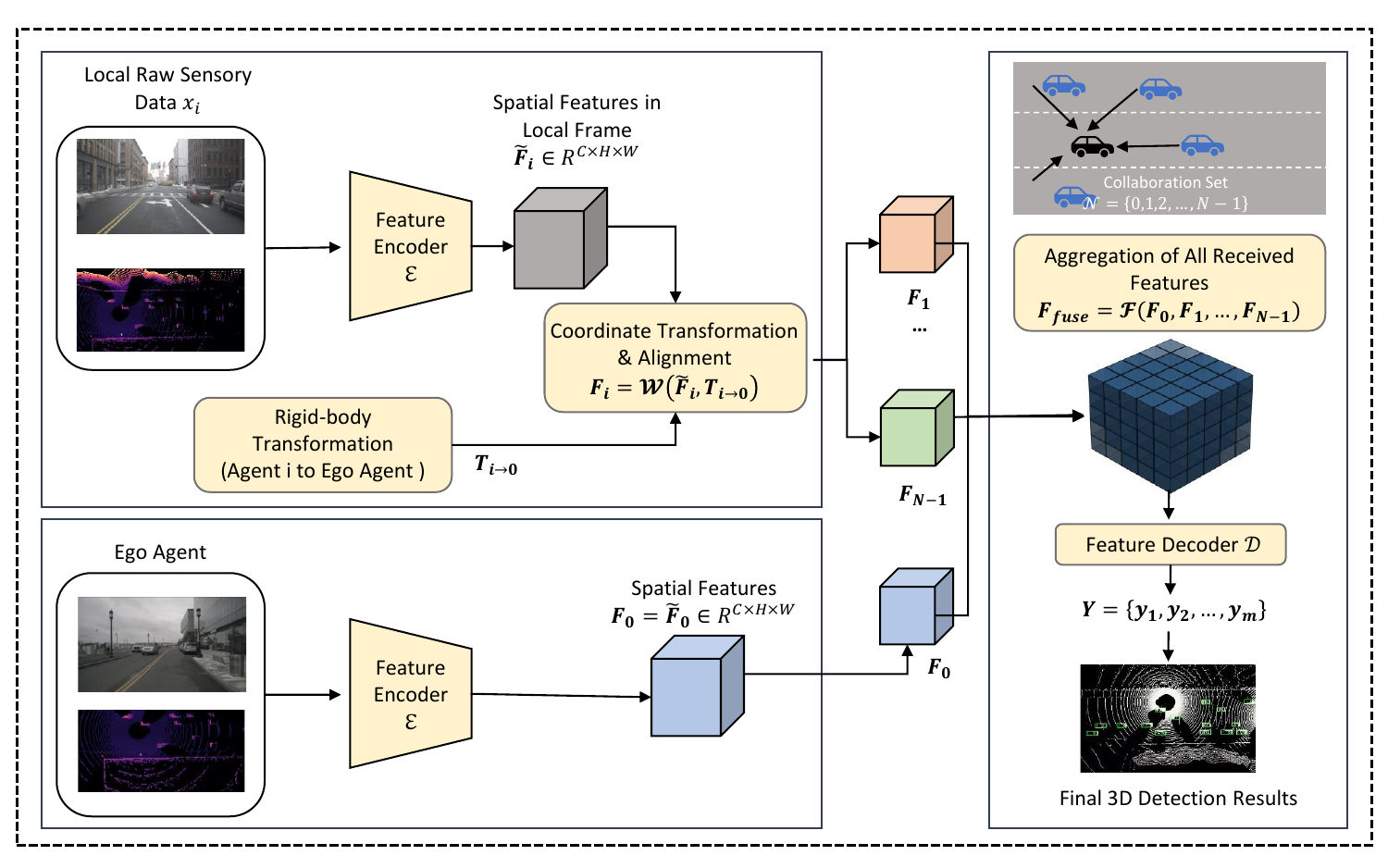}
    \caption{Illustration of the Confidence-Driven V2X CP framework.}
    \label{fig:cp_framework}
\end{figure}

We consider a collaborative network consisting of $N$ agents, denoted by $\mathcal{N}=\{0,1,\ldots,N-1\}$, where agent $0$ is the ego vehicle. The general CP pipeline is illustrated in Fig.~\ref{fig:cp_framework}. Each agent $i$ first encodes its local observation $x_i$ into a bird's-eye-view (BEV) feature representation using a feature encoder $\mathcal{E}$:
\begin{equation}
    \tilde{F}_i = \mathcal{E}(x_i), \quad \tilde{F}_i \in \mathbb{R}^{C\times H\times W},
\end{equation}
where $C$, $H$, and $W$ denote the channel, height, and width dimensions, respectively. To enable feature aggregation at the ego vehicle, the feature map of each collaborator is transformed into the ego coordinate system. Let $T_{i\rightarrow 0}$ denote the rigid-body transformation from agent $i$ to the ego agent. The aligned feature can be written as $F_i = \mathcal{W}(\tilde{F}_i,T_{i\rightarrow 0})$, where $\mathcal{W}(\cdot)$ is the spatial warping operator. The ego agent then aggregates its local feature and the received aligned features through a fusion module $\mathcal{F}$, and the fused representation is decoded by a detection head $\mathcal{D}$:
\begin{equation}
    Y_0 = \mathcal{D}\big( F_{\mathrm{fuse}}^0 \big), \quad
    F_{\mathrm{fuse}}^0 = \mathcal{F}(F_0,F_1,\ldots,F_{N-1}),
\end{equation}
where $Y_0=\{y_1,y_2,\ldots,y_M\}$ denotes the final 3D detection results of the ego vehicle, and each $y_m$ contains a 3D bounding box, category label, and confidence score.

The above formulation assumes that all intermediate features are available for fusion. In practical V2X systems, however, transmitting full BEV feature maps from all collaborators may introduce considerable communication overhead, especially when the number of agents increases or the available bandwidth is limited. To improve communication efficiency, Where2comm~\cite{hu2022where2comm} introduces a confidence-driven sparse communication mechanism that selects only perception-critical spatial regions for transmission. For each agent $i$ and communication round $k$, a confidence generator $\mathcal{G}$ produces a spatial confidence map
\begin{equation}
    C_i^{(k)} = \mathcal{G}\big(F_i^{(k)}\big), \quad C_i^{(k)}\in\mathbb{R}^{H\times W},
\end{equation}
where a larger value of $C_i^{(k)}(x,y)$ indicates that the spatial location $(x,y)$ is more informative for downstream detection. Based on the confidence map and the request map from the receiver, a binary communication mask is generated as
\begin{equation}
M_{i \to j}^{(k)} =
\begin{cases}
\Phi_{\mathrm{select}}\big(C_i^{(k)}\big), & k=0, \\[0.6em]
\Phi_{\mathrm{select}}\big(C_i^{(k)}\odot R_j^{(k-1)}\big), & k>0,
\end{cases}
\label{eq:mask_generate}
\end{equation}
where $\Phi_{\mathrm{select}}(\cdot)$ denotes the top-$p$ spatial selection function under a given communication budget, $R_j^{(k-1)}$ is the request map generated by receiver $j$ in the previous round, and $\odot$ denotes element-wise multiplication. The message transmitted from agent $i$ to agent $j$ is therefore a sparse feature map:
\begin{equation}
    Z_{i\to j}^{(k)} = M_{i\to j}^{(k)} \odot F_i^{(k)} .
    \label{eq:message_pass}
\end{equation}

After message passing, gents perform confidence-aware multi-head attention to compute fusion weights. For receiver $i$, the attention weight assigned to the message from collaborator $j$ is computed as
\begin{equation}
W_{j\to i}^{(k)} = \mathrm{MHA}_w\big(F_i^{(k)}, Z_{j\to i}^{(k)}, Z_{j\to i}^{(k)}\big) \odot C_j^{(k)},
\label{eq:attention_weight}
\end{equation}
where $\mathrm{MHA}_w(\cdot)$ denotes the multi-head attention module for estimating spatial fusion weights. The resulting feature representation is obtained by aggregating the weighted sparse messages from the communication neighbors:
\begin{equation}
F_{\mathrm{fuse}}^{i,(k)} = \mathrm{FFN}\left(\sum_{j\in \mathcal{N}_i\cup\{i\}} W_{j\to i}^{(k)} \odot Z_{j\to i}^{(k)}\right),
\label{eq:where2comm_fuse}
\end{equation}
where $\mathcal{N}_i$ is the set of collaborators connected to agent $i$, and $\mathrm{FFN}(\cdot)$ denotes a feed-forward network.

Although confidence-driven sparse communication improves efficiency, it also changes the security properties of feature-level CP. First, high-confidence or requested regions are more likely to be transmitted; therefore, if an attacker can craft malicious features that appear informative in these regions, the communication policy may unintentionally prioritize the adversarial content. Second, sparse transmission reduces the amount of redundant information available for cross-agent consistency checking, making it more difficult to distinguish benign complementary observations from carefully manipulated features. Third, attention-based fusion may assign large weights to malicious messages that align with confidence priors, causing small perturbations to have a disproportionately large influence on the final fused representation. These properties motivate the threat model and attack analysis presented below.

\subsection{Threat Model and Attack Objective}
\label{sec:threat_model}
We study feature-injection attacks against confidence-driven CP systems. The ego vehicle is assumed to be benign, while one or more collaborators may be compromised. Let $\mathcal{A}\subseteq\mathcal{N}\setminus\{0\}$ denote the set of malicious agents. A compromised agent $a\in\mathcal{A}$ can manipulate the intermediate feature transmitted to the ego vehicle by replacing the legitimate aligned feature $F_a$ with an adversarial feature $\tilde{F}_a$. This setting follows the common CP security assumption that the attacker has already compromised a collaborator or its perception software stack, while the ego-side fusion and detection modules remain unchanged.

The attacker is considered under the following capabilities and constraints:
\begin{itemize}
    \item \textbf{Feature manipulation capability:} The attacker can modify its own transmitted intermediate feature after local encoding and coordinate alignment, i.e., $\tilde{F}_a=F_a+\delta_a$, where $\delta_a$ is the adversarial perturbation.
    \item \textbf{Model knowledge:} The attacker has white-box access to the deployed CP model, including the encoder, communication module, fusion module, and detection head. This allows end-to-end gradient-based optimization of adversarial perturbations.
    \item \textbf{Limited cross-agent observability:} The attacker cannot directly access the private intermediate features of benign collaborators. It can use its own local feature, the public communication protocol, and ego-related information available during collaboration, such as the ego feature or confidence map.
    \item \textbf{Perturbation constraint:} To avoid trivially detectable feature corruption, the perturbation is bounded by an $\ell_p$ norm constraint, i.e., $\|\delta_a\|_p\leq\epsilon$.
\end{itemize}

The attack objective is to degrade the ego vehicle's perception results by inducing false negatives, false positives, or inaccurate bounding boxes after feature fusion. For a single attacker $a$, the adversarial feature is $\tilde{F}_a = F_a + \delta_a, \quad \|\delta_a\|_p \leq \epsilon$. The ego-side prediction under attack is formulated as
\begin{equation}
    Y_0^{\delta} = \mathcal{D}\left(\mathcal{F}\left(F_0,F_1,\ldots,\tilde{F}_a,\ldots,F_{N-1}\right)\right).
\end{equation}
The attacker seeks to maximize a detection-oriented adversarial loss $\max_{\|\delta_a\|_p\leq\epsilon} \; \mathcal{L}_{\mathrm{adv}}\left(Y_0^{\delta},Y_0^{\mathrm{gt}}\right)$,
where $Y_0^{\mathrm{gt}}$ denotes the ground-truth 3D annotations of the ego vehicle, and $\mathcal{L}_{\mathrm{adv}}(\cdot)$ measures the attack effect on downstream detection.

In the multi-attacker setting, each malicious agent independently generates its own perturbation and transmits an adversarial feature to the ego vehicle. The attack objective can be generalized as
\begin{equation}
\begin{aligned}
    &\max_{\{\delta_a\}_{a\in\mathcal{A}}} \; \mathcal{L}_{\mathrm{adv}}\left(Y_0^{\delta},Y_0^{\mathrm{gt}}\right), \\
    &\mathrm{s.t.}\quad \tilde{F}_a=F_a+\delta_a, \quad \|\delta_a\|_p\leq\epsilon, \quad \forall a\in\mathcal{A},
\end{aligned}
\label{eq:multi_attack_objective}
\end{equation}
where
\begin{equation}
    Y_0^{\delta}=\mathcal{D}\left(\mathcal{F}\left(F_0,\{F_i\}_{i\notin\mathcal{A}},\{\tilde{F}_a\}_{a\in\mathcal{A}}\right)\right).
\end{equation}
Unless otherwise specified, the multi-attacker setting considered in this paper assumes independent attackers rather than explicitly coordinated attackers. This assumption is consistent with the experimental design in Section~VI and allows us to evaluate whether a defense can remain reliable when several malicious agents simultaneously inject benign-looking perturbations. Securing the underlying V2X communication channel, authentication protocol, and message integrity is also important in practice~\cite{zhao2023generic}, but it is orthogonal to the feature-level robustness problem studied in this work.

\section{PB Attack Against Confidence-Driven CP}

\begin{figure*}[t]
    \centering
    \includegraphics[width=0.92\textwidth]{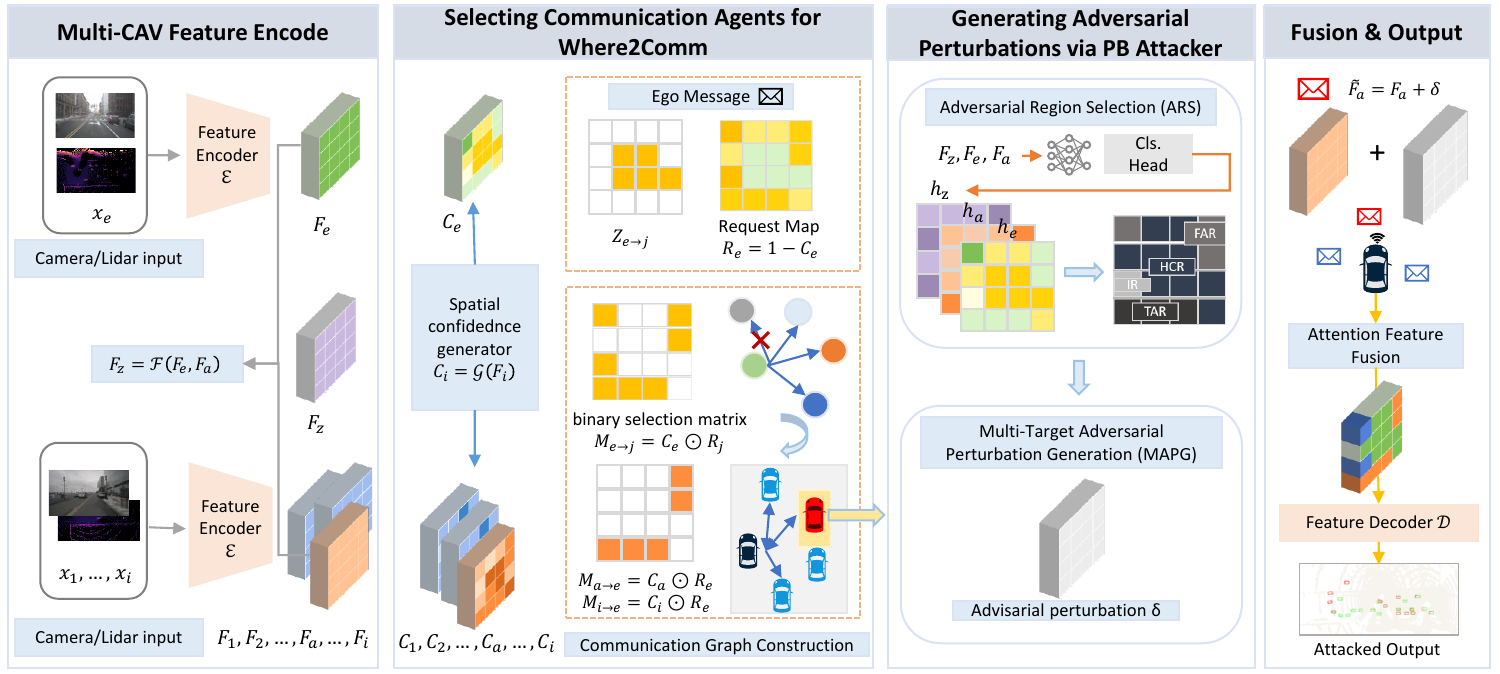}
    \caption{Illustration of the PB attack against Confidence-Driven CP. PB exploits confidence-aware communication by selecting attack-sensitive regions through ARS and generating region-specific perturbations through MAPG.}
    \label{fig:pb_where2comm}
\end{figure*}

This section analyzes why the PB attack~\cite{lin2025pretend} is particularly effective against confidence-driven CP frameworks such as Where2comm. PB is an existing stealthy feature-injection attack rather than a method proposed in this paper. We use it as a representative threat to expose the security weakness of confidence-based communication and attention-based fusion. Unlike PGD~\cite{madry2017towards} or BIM~\cite{kurakin2018adversarial}, which generally optimize perturbations over the entire feature map, PB explicitly considers the spatial semantics of different BEV regions. Its goal is twofold: degrading the ego vehicle's final detection results and preserving benign-like feature characteristics so that the malicious collaborator is difficult to identify by simple similarity-based defenses.

As shown in Fig.~\ref{fig:pb_where2comm}, PB contains two major modules: Attack Region Selection (ARS) and Multi-target Adversarial Perturbation Generation (MAPG). ARS partitions the feature map into several functional regions according to the confidence responses of the ego vehicle, the attacker, and the fused perception output. MAPG then assigns different optimization objectives to these regions, allowing the attacker to selectively suppress real objects, introduce false detections, and maintain partial feature consistency. This region-aware design matches the communication logic of Where2comm and enables malicious features to pass through the sparse communication mask and receive non-negligible attention during fusion.

\subsection{Attack Region Selection}
PB first uses the classification head of the deployed 3D detector to obtain three confidence heatmaps: the ego confidence heatmap $h_{\mathrm{e}}$, the attacker confidence heatmap $h_{\mathrm{a}}$, and the fused confidence heatmap $h_{\mathrm{z}}$. These heatmaps describe how different spatial locations are perceived before and after collaboration. Given two confidence thresholds $\mu_h$ and $\mu_l$, ARS partitions the BEV feature space into four functional regions:
\begin{equation}
    \{\mathcal{R}_{\mathrm{HCR}},\mathcal{R}_{\mathrm{TAR}},\mathcal{R}_{\mathrm{FAR}},\mathcal{R}_{\mathrm{IR}}\}
    = \mathrm{ARS}(h_{\mathrm{e}},h_{\mathrm{a}},h_{\mathrm{z}};\mu_h,\mu_l),
\label{eq:ars_partition}
\end{equation}
where $\mathcal{R}_{\mathrm{HCR}}$, $\mathcal{R}_{\mathrm{TAR}}$, $\mathcal{R}_{\mathrm{FAR}}$, and $\mathcal{R}_{\mathrm{IR}}$ denote the High Confidence Region, True Ambiguous Region, False Ambiguous Region, and Imperceptible Region, respectively.

The High Confidence Region contains locations with strong and relatively consistent object responses. Directly corrupting these locations may produce obvious feature deviations and make the malicious agent easier to detect. Therefore, PB tends to preserve partial consistency in this region. The True Ambiguous Region corresponds to locations where real objects are uncertain or weakly perceived by the ego vehicle; suppressing confidence in this region can cause missed detections. The False Ambiguous Region and Imperceptible Region are exploited to create false object responses in locations where the ego vehicle has insufficient evidence. In this way, PB avoids a uniform perturbation pattern and instead allocates perturbations according to the semantic role of each spatial region.

This region-selection strategy directly exploits confidence-driven communication. Since Where2comm selects high-value spatial locations according to confidence maps and request maps, adversarial features that appear informative in ambiguous or complementary regions are more likely to be included in the transmitted sparse message $Z_{a\to 0}$. Consequently, the malicious feature is not necessarily filtered out by the communication policy. Instead, it may be treated as useful complementary information and propagated to the ego-side fusion module.

\subsection{Multi-target Adversarial Perturbation Generation}
After region partitioning, MAPG generates perturbations with region-specific objectives. In the High Confidence Region, PB introduces a regularization term to maintain detection quality and preserve a benign-looking feature response. In the True Ambiguous Region, PB suppresses confidence scores associated with real objects, encouraging false negatives. In the False Ambiguous Region and Imperceptible Region, PB increases confidence scores to induce false positives. The overall adversarial objective is formulated as
\begin{align}
\mathcal{L}_{\mathrm{adv}} 
&= \sum_{i \in \mathrm{HCR}} \left[ \log(1-y_i^s) \cdot \mathrm{IoU}(y_i',y_i) - \lambda R(y_i',y_i) \right] \notag \\
& + \sum_{i \in \mathrm{TAR}} \log(1-y_i^s) \cdot \mathrm{IoU}(y_i',y_i)  + \sum_{i \in \mathrm{FAR}\cup\mathrm{IR}} \log(y_i^s),
\label{eq:pb_loss}
\end{align}
where $y_i$ denotes an original detection result, $y_i'$ denotes the corresponding post-attack bounding box, $y_i^s$ is the post-attack confidence score, and $\mathrm{IoU}(y_i',y_i)$ measures the geometric overlap between the perturbed and original bounding boxes. The regularization term $R(y_i',y_i)$ is used in the High Confidence Region to avoid excessive corruption of highly reliable detections, and $\lambda$ controls its strength.

The perturbation is iteratively updated in the feature space by maximizing $\mathcal{L}_{\mathrm{adv}}$ under the norm constraint in Section~\ref{sec:threat_model}. A typical projected gradient update can be written as
\begin{equation}
    \delta_a^{t+1} = \Pi_{\|\delta\|_p\leq\epsilon}\left(\delta_a^t + \alpha \cdot \mathrm{sign}\left(\nabla_{\delta_a}\mathcal{L}_{\mathrm{adv}}\right)\right),
\label{eq:pb_update}
\end{equation}
where $\alpha$ is the step size and $\Pi_{\|\delta\|_p\leq\epsilon}(\cdot)$ denotes projection onto the perturbation budget. The malicious feature transmitted by attacker $a$ is then given by
\begin{equation}
    \tilde{F}_a = F_a + \delta_a.
\end{equation}

\subsection{Why PB Is Amplified by Where2comm}
The effectiveness of PB comes from the interaction between its region-aware perturbation design and the confidence-driven CP pipeline. First, PB places adversarial effects in regions that can be interpreted as informative or complementary by the communication module. These regions are more likely to survive the sparse selection mask in Eq.~\eqref{eq:mask_generate}. Second, PB maintains partial consistency in high-confidence regions, which reduces the chance of being detected by defenses that rely only on global feature distance or single-level similarity. Third, after the malicious sparse message is transmitted, the attention-based fusion module in Eq.~\eqref{eq:attention_weight} may assign non-trivial weights to the adversarial regions because they are aligned with confidence priors. As a result, the injected perturbations can be amplified during feature fusion and ultimately distort the ego vehicle's final detection results.

\begin{figure*}[t]
    \centering
  \includegraphics[width=0.87\textwidth]{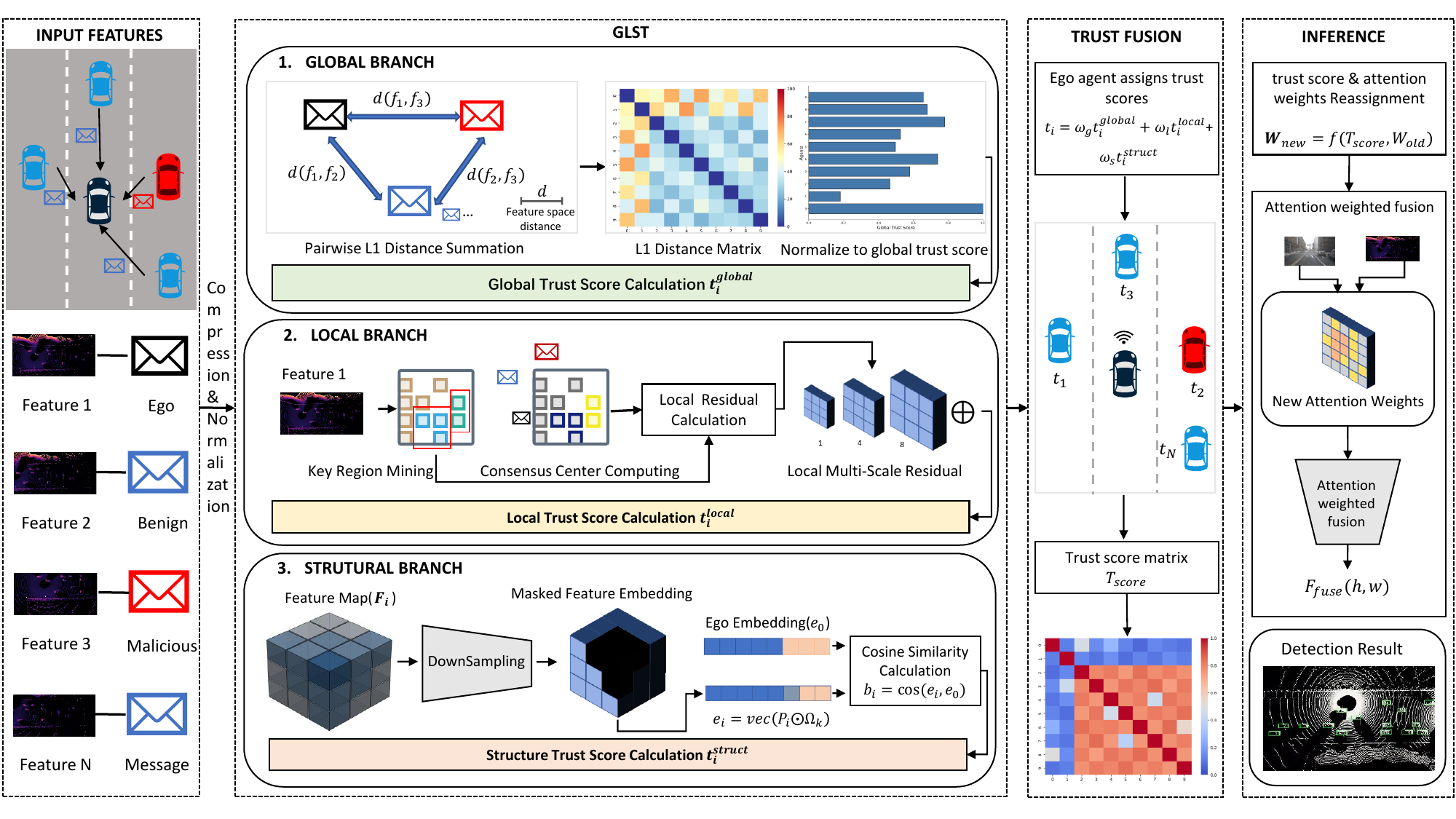}
    \caption{Overview of the proposed GLST defense framework. GLST estimates collaborator reliability from global feature consistency, critical-region local residual consistency, and ego-guided structural consistency, and then uses the fused trust scores to modulate the attention-based feature fusion process}
    \label{fig:framework}
\end{figure*}

This analysis also explains why PB is a suitable attack for evaluating robust CP defenses. A defense that only checks global feature consistency may fail when the attacker preserves benign-like statistical characteristics, while a defense that only examines local residuals may be unstable when several attackers create a pseudo-consensus. Therefore, an effective defense should evaluate collaborator reliability from multiple complementary perspectives. This observation motivates the proposed GLST framework in Section~V, which combines global consistency, local robust residuals, and structural consistency to suppress stealthy feature injection under both single-attacker and multi-attacker settings.


\section{Proposed Defense Framework: GLST}
\subsection{Framework Overview}

Our analysis reveals a fundamental limitation of defenses that rely solely on a single global statistical measure,  are vulnerable in multi-attacker scenarios due to consensus bias.  When there are multiple attackers in the system,  their small inter-attacker distances can artificially increase their individual trust scores, 
which in turn systematically shifts the fusion center and degrades the overall robustness of the system. 

To address this issue, we propose the GLST defense framework. The core idea is to extend trust evaluation from a single global signal to a joint modeling of three complementary perspectives: global consistency provides a coarse-grained prior, local residual captures fine-grained anomalies in critical regions,  and structural consistency enforces the stability of semantic topology. These three signals are combined through a weighted summation strategy to avoid trust collapse caused by multiplicative coupling effects. As illustrated in Fig.~\ref{fig:framework}, GLST consists of four major stages.  Given the intermediate BEV feature $F_i$ from each agent $i$, GLST first computes a global statistical trust score by measuring the deviation of each agent from the collective feature distribution. Then, guided by the ego confidence and feature response, GLST selects critical spatial regions and estimates multi-scale local residual trust. Next, an ego-referenced structural consistency score is computed to capture abnormal spatial semantic patterns. Finally, the three trust signals are aggregated through a weighted summation strategy and injected into the feature fusion process as a trust-aware reweighting factor. This design enables soft suppression of suspicious agents while preserving useful complementary information from benign collaborators.


\subsection{Global Feature Consistency Trust}
\label{subsec:global_trust}

The global consistency branch aims to provide a coarse-grained reliability prior for each collaborative agent. The intuition is that benign agents observing the same driving scene should produce feature representations that are statistically compatible with the majority distribution,  whereas malicious agents tend to introduce abnormal feature patterns that deviate from the collective representation.  Therefore, GLST first evaluates agent reliability by measuring the global feature discrepancy among all agents.

Specifically, the spatial feature map $F_i$ of each agent is first downsampled to reduce computational cost and then normalized to remove scale variation:
\begin{equation}
\bar{F}_i = \mathrm{Norm}\big(\mathrm{Pool}(F_i,k)\big),
\label{eq:global_pool_norm}
\end{equation}
where $\mathrm{Pool}(\cdot,k)$ denotes spatial pooling with kernel size $k$, and $\mathrm{Norm}(\cdot)$ represents $\ell_2$ normalization. The pooling operation reduces computational overhead, while normalization suppresses the influence of feature magnitude and makes the subsequent comparison focus more on distribution-level differences.

Based on the normalized feature representation,  the cumulative deviation score of agent $i$ by measuring its pairwise $\ell_1$ distance to the other agents can be computed as
\begin{equation}
d_i = \frac{1}{N-1} \sum_{j \in \mathcal{N}, j \neq i} \left\| \bar{F}_i - \bar{F}_j \right\|_1, \quad i \in \mathcal{N}.
\label{eq:feature_distance}
\end{equation}
A larger $d_i$ indicates that the feature of agent $i$ is farther away from the collective distribution and should therefore be assigned a lower trust score.

Since the ego agent provides the reference view for feature fusion and should not be mistakenly suppressed, we explicitly protect the ego feature by assigning its deviation score to the minimum deviation among the collaborative agents:
\begin{equation}
d_0 = \min_{i \in \mathcal{N}} d_i .
\label{eq:ego_deviation_protection}
\end{equation}
The deviation scores are then normalized and converted into global trust scores:
\begin{equation}
t_i^{\mathrm{global}} = 1 - \frac{d_i}{\max_{j \in \mathcal{N}} d_j + \varepsilon},\quad i \in \mathcal{N},
\label{eq:t_global}
\end{equation}
where $\varepsilon$ is a small constant used to avoid numerical instability.

The global trust score serves as the first reliability prior in GLST. It can effectively suppress agents that are globally inconsistent with the majority feature distribution. 
However, global statistical consistency alone is insufficient in coordinated attack scenarios.  When multiple attackers collude and generate mutually similar malicious features, their distances to each other may reduce the cumulative deviation score and cause overestimated trust values. Therefore, GLST further introduces local residual trust and structural consistency trust to compensate for the weakness of purely global estimation.

\subsection{Local Robust Consensus Residual Module}

Although global consistency provides an efficient reliability prior, it treats the entire feature map uniformly and is therefore insensitive to localized manipulations. In collaborative perception, adversarial perturbations often concentrate on semantically important regions, such as object boundaries, foreground areas, or locations with high detection uncertainty. These regions occupy only a small portion of the BEV feature map but can significantly influence the final detection result. To capture such fine-grained anomalies, GLST introduces a local robust consensus residual module.

\textbf{Critical Region Selection.} Before computing local residuals, GLST first identifies a set of critical spatial regions $\Omega$ from the ego feature. 
The ego feature is used because it is directly available at the fusion center and provides a stable reference for selecting perception-sensitive regions. 
Two complementary spatial cues are extracted from the ego agent.

We first compute the uncertainty map $U \in \mathbb{R}^{H \times W}$ from the ego detection confidence as follows
\begin{equation}
U(x,y) = 1 - \max_{k} 
\mathrm{softmax}\left( \Phi_{\mathrm{ds}}(F_0) \right)_k ,
\label{eq:uncertainty}
\end{equation}
where $\Phi_{\mathrm{ds}}(\cdot)$ denotes the detection score prediction head, and $k$ indexes the object class. 
A larger value of $U(x,y)$ indicates lower prediction confidence and thus higher uncertainty.

Then, the feature response intensity map $S \in \mathbb{R}^{H \times W}$ can be determined by
\begin{equation}
S(x,y) = 
\frac{
\left\| F_0(:,x,y) \right\|_2
}{
\max_{x,y} \left\| F_0(:,x,y) \right\|_2 + \epsilon
}.
\label{eq:spatial_confidence}
\end{equation}
This map highlights locations with strong feature activations, which are more likely to correspond to foreground objects or perception-relevant regions.

The final critical-region mask is obtained by selecting locations that simultaneously exhibit high uncertainty and strong feature responses, which is given by
\begin{equation}
\Omega =
\left\{
(x,y) \mid
U(x,y) \geq \mathcal{Q}_{1-p}(U)
\;\land\;
S(x,y) \geq \mathcal{Q}_{1-p}(S)
\right\},
\label{eq:omega_set}
\end{equation}
where $\mathcal{Q}_{1-p}(\cdot)$ denotes the $(1-p)$-quantile, and $p$ controls the proportion of selected regions. This intersection strategy suppresses irrelevant background areas and forces the local trust evaluation to focus on regions that are both uncertain and semantically active.

\textbf{Multi-Scale Consensus Residual.} After obtaining the critical-region mask, GLST estimates the local deviation of each agent from a robust consensus center. 
For each spatial scale $s \in \mathcal{S}$, the feature of agent $i$ is downsampled as $F_i^{(s)}$, and the corresponding mask is denoted as $\Omega^{(s)}$. 
To reduce the influence of suspicious agents during consensus construction, we use the global trust score \eqref{eq:t_global} as a reliability weight, and the consensus center is formulated as
\begin{equation}
C^{(s)} =
\begin{cases}
F_0^{(s)}, & N \leq 3, \\ 
\sum_{i} w_i F_i^{(s)}, w_i =\dfrac{t_i^{\mathrm{global}}}{\sum_{j} t_j^{\mathrm{global}} + \epsilon},& N > 3.
\end{cases}
\label{eq:C_s}
\end{equation}
When the number of agents is small, the ego feature is directly adopted as the consensus center to avoid unstable estimation caused by insufficient collaborators. When more agents are available, a globally weighted consensus center is used to reduce the contribution of potentially malicious agents.

Given the consensus center \eqref{eq:C_s}, the residual map of agent $i$ within the critical region is computed as
\begin{equation}
R_i^{(s)}(x,y) =
\left\|
F_i^{(s)}(:,x,y) - C^{(s)}(:,x,y)
\right\|_2 .
\label{eq:R_i_s}
\end{equation}
The local residual score at this scale is then defined as a combination of the average residual and the high-percentile residual within the critical region:
\begin{equation}
a_i^{(s)} =0.7 \cdot \mathbb{E}_{(x,y) \in \Omega^{(s)}} \left[ R_i^{(s)}(x,y) \right]+0.3 \cdot \mathcal{Q}_{0.9}\left( R_i^{(s)} \mid \Omega^{(s)} \right).
\label{eq:a_i_s}
\end{equation}
The mean term $\mathbb{E}(\cdot)$ captures the overall local inconsistency, while the 90th-percentile term $Q_{0.9}(\cdot)$  is introduced to capture extreme deviations concentrated on a small number of spatial locations. Even when attackers successfully conceal the average residual magnitude, these localized abnormal responses can still expose malicious manipulations.

The residual scores of agent $i$ from multiple spatial scales are aggregated as $\bar{a}_i =\frac{1}{|\mathcal{S}|}\sum_{s \in \mathcal{S}} a_i^{(s)}$. Finally, the local residual score is converted into a local trust score through exponential decay:
\begin{equation}
t_i^{\mathrm{local}} =\exp \left(-\beta_l\cdot\frac{\bar{a}_i}{\max_{j \in \mathcal{N}} \bar{a}_j }\right),
\label{eq:t_i_local}
\end{equation}
where $\beta_l$ controls the sensitivity of local trust to residual deviations. Agents that consistently deviate from the consensus center across multiple scales receive lower local trust scores. Since it is difficult for attackers to maintain consistent camouflage simultaneously at different spatial resolutions, the multi-scale design improves robustness against localized and coordinated feature manipulations.

\subsection{Structural Consistency Check}

Some stealthy attacks can mimic benign agents in terms of global statistics and even maintain small local residual magnitudes. However, malicious feature manipulation may still distort the internal spatial structure of the feature map, such as activation patterns in object regions, edge structures, and spatial correlations across feature channels. To capture these higher-level spatial discrepancies, GLST introduces an ego-referenced structural consistency branch.

Different from the local residual branch, which measures point-wise deviation from a consensus center, the structural consistency branch evaluates whether the overall spatial semantic pattern of each agent is compatible with the ego agent. Specifically, each agent feature is first spatially pooled and then masked by the downsampled critical-region mask:
\begin{equation}
P_i = \mathrm{AvgPool}_d(F_i),
\quad
e_i = \mathrm{vec}\left( P_i \odot \Omega^{(d)} \right),
\label{eq:pooling_vec}
\end{equation}
where $\Omega^{(d)}$ denotes the downsampled mask, $\odot$ represents element-wise multiplication, and $\mathrm{vec}(\cdot)$ represents the flattening operation.

The structural similarity between agent $i$ and the ego agent $e_0$ is measured by cosine similarity
\begin{equation}
b_i =\cos(e_i, e_0)=\frac{e_i^{T} e_0}{\|e_i\|_2 \|e_0\|_2 + \epsilon}.
\label{eq:struct_cosine}
\end{equation}
A larger $b_i$ indicates that agent $i$ preserves a spatial semantic structure more similar to the ego reference. Based on \eqref{eq:struct_cosine}, the structural dissimilarity is then normalized and transformed into a trust score, which is given by
\begin{equation}
t_i^{\mathrm{struct}} =\exp \left(-\beta_s\cdot\frac{(1-b_i)/2}{\max_{j \in \mathcal{N}} (1-b_j)/2 + \epsilon}\right),
\label{eq:t_struct}
\end{equation}
where $\beta_s$ controls the sensitivity of structural trust. 

This branch provides an additional reliability criterion that is independent of global distribution distance and local residual magnitude. 
Therefore, even when attackers partially bypass statistical consistency checks, 
their abnormal spatial semantic structures can still be detected by the structural consistency branch.

\subsection{Trust-Aware Feature Re-weighting and Fusion}

After obtaining global, local, and structural trust scores, GLST aggregates them into a unified agent reliability score. Instead of using multiplicative coupling, we adopt weighted summation to compute the final trust score which is
\begin{equation}
t_i =\omega_g t_i^{\mathrm{global}}+\omega_l t_i^{\mathrm{local}}+\omega_s t_i^{\mathrm{struct}},
\label{eq:combined_t}
\end{equation}
where $\omega_g$, $\omega_l$, and $\omega_s$ denote the weights of the three trust branches, with $\omega_g+\omega_l+\omega_s=1$ and $\omega_g,\omega_l,\omega_s \geq 0$.

The reason for using weighted summation is twofold. 
First, multiplicative aggregation may over-penalize benign agents when one trust branch produces an unstable low value, 
which can increase false-positive suppression. 
Second, weighted summation allows different trust cues to compensate for each other under different attack patterns. 
For example, when coordinated attackers weaken global deviation, local and structural trust can still provide effective discrimination. 
Similarly, when local residuals are affected by perception noise, global consistency and structural similarity can stabilize the final decision.

%

To incorporate the trust score into the fusion process, the final trust \eqref{eq:combined_t} are then normalized across agents as
\begin{equation}
\hat{t}_i = \frac{t_i}{\sum_{j \in N} t_j + \epsilon}
\label{eq:normalized_t}
\end{equation}
Then, injecting the normalized trust score \eqref{eq:normalized_t} into the original attention distribution, trust-aware modulation is then performed as
\begin{equation}
\tilde{W}_{i \to 0}(h,w) = \frac{\hat{t}_i W_{i \to 0}(h,w)}{\sum_{j \in N} \hat{t}_j W_{j \to 0}(h,w) + \epsilon}
\label{eq:weight_fusion}
\end{equation}
where $W_{i \to 0}(h,w)$ denote the original attention weight from agent $i$ to the ego agent at BEV location $(h,w)$. The final fusion process can therefore be expressed as
\begin{equation}
F_{\mathrm{fuse}}(h,w) = \mathrm{FFN}\left( \sum_{i \in N} \tilde{W}_{i \to 0}(h,w) Z_{i \to 0}(h,w) \right)
\label{eq:feature_fusion}
\end{equation}
where $Z_{i \to 0}(h,w)$ denotes the transformed feature transmitted from agent $i$ to the ego coordinate system, and $\mathrm{FFN}(\cdot)$ denotes the feed-forward fusion module.

This trust-aware fusion mechanism can be regarded as a soft gating operation over the original attention weights. 
Agents with high reliability are assigned larger effective fusion weights, 
whereas suspicious agents are continuously suppressed before they contaminate the fused BEV representation. 
Compared with hard filtering, this soft reweighting strategy avoids discarding potentially useful observations from partially reliable agents, 
thereby improving the robustness and stability of collaborative perception under complex attack scenarios.



\section{Experiments}

\subsection{Experimental Setup}
\textbf{Datasets and Detector.} We evaluate the vulnerability of Where2comm-based collaborative perception under the existing PB attack~\cite{lin2025pretend} and assess the effectiveness of the proposed GLST defense framework on the widely used OPV2V dataset~\cite{xu2022opv2v}, a large-scale open-source benchmark for V2V collaborative perception. OPV2V is built upon the OpenCDA~\cite{xu2021opencda} co-simulation framework and the CARLA simulator~\cite{dosovitskiy2017carla}. It contains $70$ diverse driving scenarios collected from eight towns, including $11,464$ LiDAR frames and 232,913 annotated 3D vehicle bounding boxes. Each frame contains 2--7 CAVs, and each CAV is equipped with a 64-beam LiDAR sensor with complete pose transformations, point-cloud observations, and 3D annotations. All experiments are conducted based on the OpenCOOD framework~\cite{xu2022opv2v}.
We adopt PointPillars~\cite{lang2019pointpillars} as the detector backbone and use the confidence-aware multi-head attention fusion module of Where2comm as the CP framework.

\textbf{Attack Details.} We evaluate three feature-level adversarial attacks against Where2comm, including PGD~\cite{madry2017towards}, BIM~\cite{kurakin2018adversarial}, and PB~\cite{lin2025pretend}. For PB, we follow the iterative sign-gradient optimization setting with $T=10$ iterations, learning rate $l_r=0.1$, and regularization weight $\lambda=5.0$. The region partition thresholds are set to $0.7$ and $0.5$, respectively. For the baseline attacks, PGD uses random initialization, while BIM uses zero initialization. Both PGD and BIM use the same number of iterations and learning rate as PB, with the perturbation budget set to $\epsilon=0.3$. All attackers independently optimize their perturbations using the same detection loss objective.

\textbf{Defense Details.} We compare GLST with two representative CP defense methods, ROBOSAC~\cite{li2023among} and LUCIA~\cite{wang2025threat}. ROBOSAC adopts a random sampling-based consistency strategy, and its sampling budget is set to 10. LUCIA constructs trust weights from pairwise feature $\mathcal{L}_1$ distances, with the compression ratio set to $r=32$. For GLST, the fusion weights of the global, local, and structural trust components are set to $\omega_g=0.30$, $\omega_l=0.50$, and $\omega_s=0.20$, respectively. The local residual scales are set to $s\in \{1,4,8\}$, and the critical-region ratio is set to $p=0.3$. All defense methods are evaluated under the same detector, input modality, communication setting, and attack configuration for fair comparison.

\begin{figure}[t]
    \centering
    {
    \subfigure[No Attack]
     {\includegraphics[width=0.47\linewidth]{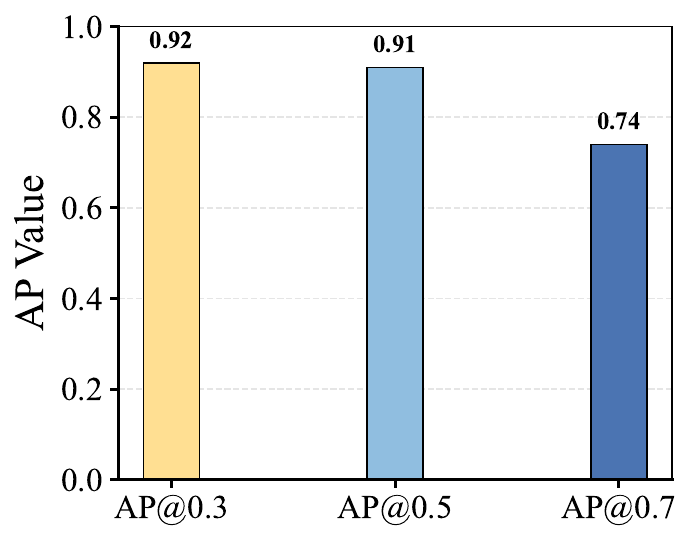} \label{fig:attack_single_No_Attack}}
     \subfigure[PGD]
   {\includegraphics[width=0.47\linewidth]{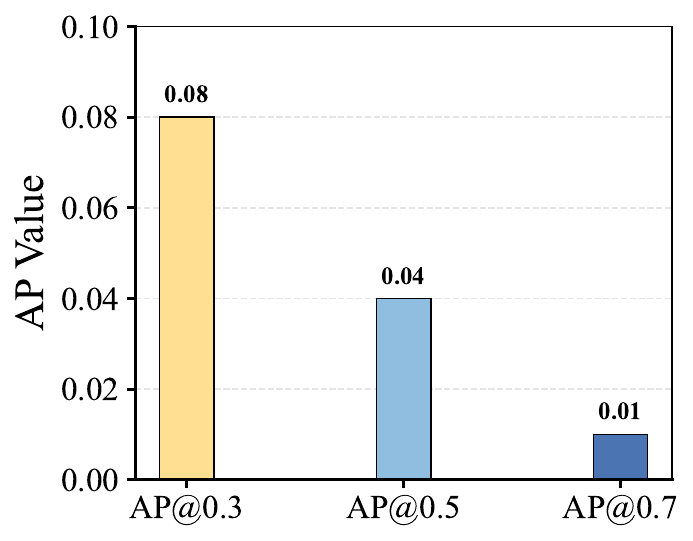} \label{fig:attack_single_PGD}}
    \subfigure[BIM]
   {\includegraphics[width=0.47\linewidth]{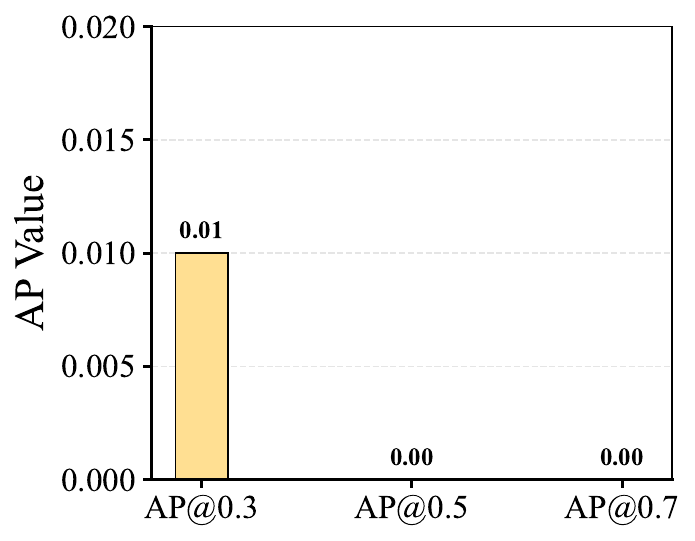} \label{fig:attack_single_BIM}}
    \subfigure[PB]
    {\includegraphics[width=0.47\linewidth]{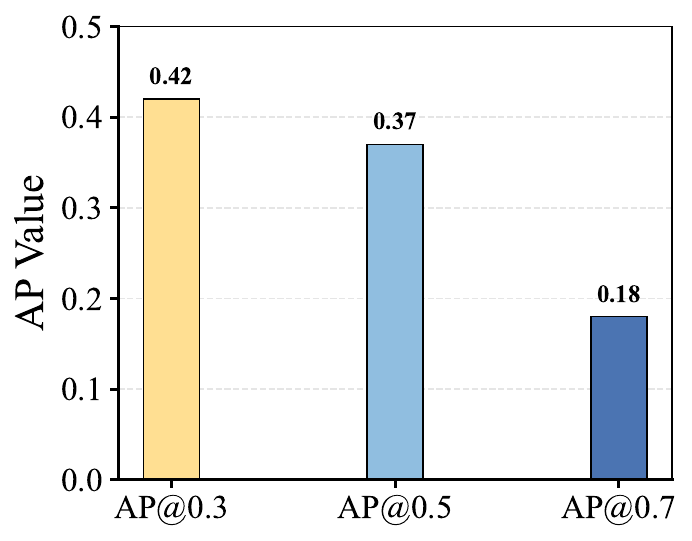} \label{fig:attack_single_PB}}
     }
 \caption{AP values at 0.3, 0.5, and 0.7 IoU thresholds after applying PGD, BIM, and PB attacks under the single-attacker setting.}
\label{fig:single_attacker_performance}
\end{figure}

\textbf{Evaluation Metrics.} We use Average Precision (AP) to evaluate 3D object detection performance under Intersection-over-Union (IoU) thresholds of $0.3$, $0.5$, and $0.7$ (denoted as AP@0.3, AP@0.5, and AP@0.7, respectively). A higher AP indicates better detection performance. In adversarial settings, AP degradation reflects the attack effectiveness, whereas AP recovery after applying a defense method indicates the defense effectiveness. To further evaluate attack stealthiness, we report the average number of false positives per frame under an IoU threshold of 0.5, which is $\mathrm{FP}_{\mathrm{IoU}0.5} = \frac{\sum_{i=0}^{\mathrm{Frame}} \mathrm{FP}_i}{\mathrm{Frame}}$, where $\mathrm{FP}_i$ denotes the number of false-positive detections in frame $i$ under an IoU threshold of 0.5, and $Frame$ is the total number of evaluated frames.
A larger $\mathrm{FP}_{\mathrm{IoU}0.5}$ indicates weaker stealthiness. Conversely, a smaller value indicates stronger visual stealthiness and greater difficulty for defense mechanisms to suppress the attack.

\subsection{Attack Performance Analysis}

To evaluate the attack effectiveness of different feature-injection strategies, we first examine the detection performance of Where2comm under the single-attacker setting, as shown in Fig.~\ref{fig:single_attacker_performance}. Without attacks, Where2comm maintains strong detection performance, achieving 0.91 AP@0.5 and 0.74 AP@0.7. After adversarial features are injected, the detection accuracy drops sharply across all IoU thresholds, demonstrating that the confidence-driven communication and attention-based fusion pipeline is highly vulnerable to malicious intermediate features. Among the evaluated attacks, PGD and BIM cause the most severe performance collapse. PGD reduces AP@0.5 to 0.04, while BIM almost completely disables the detector, reducing AP@0.5 to nearly zero. PB also substantially degrades the perception performance, reducing AP@0.5 and AP@0.7 to 0.37 and 0.18, respectively. Although PB is less destructive than PGD and BIM in terms of raw AP reduction, it still causes a significant loss of cooperative detection capability, confirming its effectiveness against confidence-driven collaborative perception.

\begin{figure}[t]
    \centering
    {
    \subfigure[No Attack]
     {\includegraphics[width=0.47\linewidth]{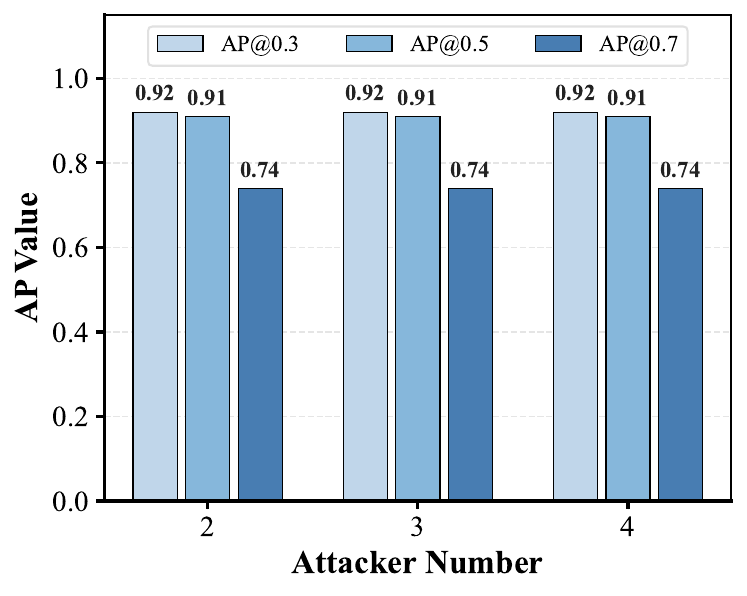} \label{fig:attack_multi_No_Attack}}
     \subfigure[PGD]
   {\includegraphics[width=0.47\linewidth]{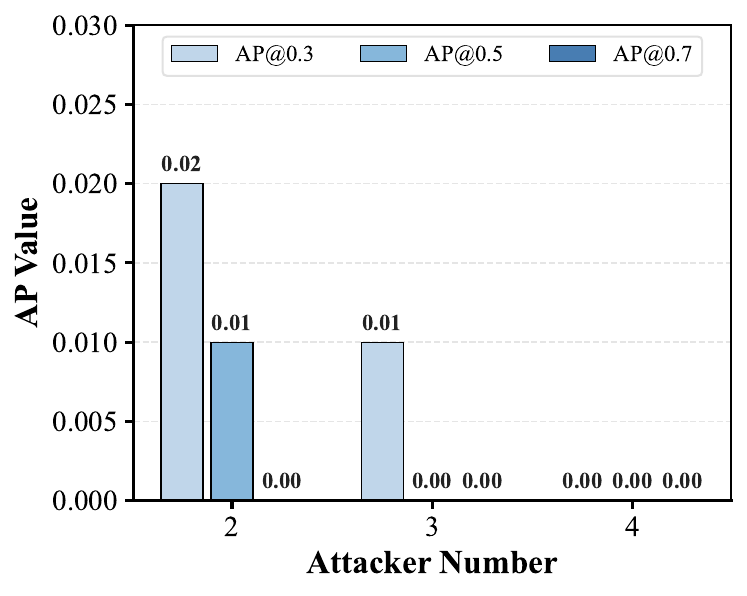}\label{fig:attack_multi_PGD}}
    \subfigure[BIM]
   {\includegraphics[width=0.47\linewidth]{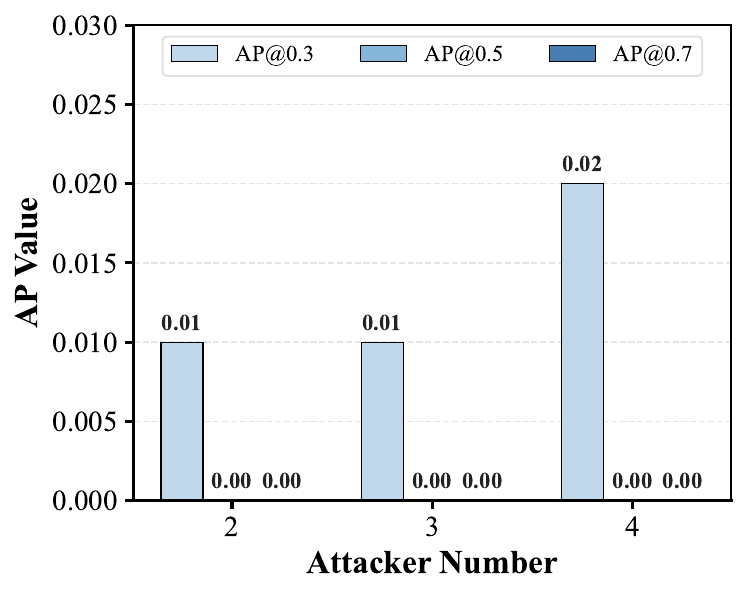} \label{fig:attack_multi_BIM}}
    \subfigure[PB]
    {\includegraphics[width=0.47\linewidth]{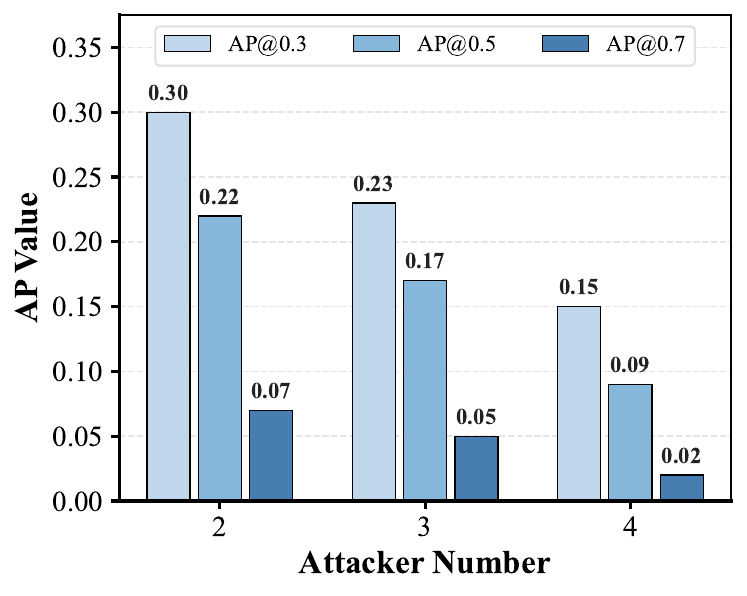} \label{fig:attack_multi_PB}}
     }
 \caption{AP values at 0.3, 0.5, and 0.7 IoU thresholds after applying PGD, BIM, and PB attacks under the multi-attacker setting.}
\label{fig:multi_attacker_performance}
\end{figure}

\begin{figure*}[t]
    \centering
    {
    \subfigure[PB attack without defence]
     {\includegraphics[width=0.32\linewidth]{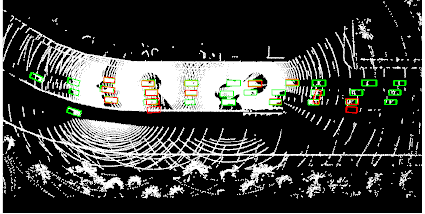} \label{fig:nodefense-PB}}
     \subfigure[PGD attack without defence]
   {\includegraphics[width=0.32\linewidth]{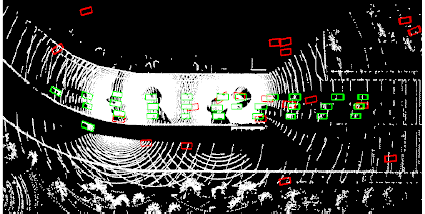}\label{fig:nodefense-PGD}}
    \subfigure[BIM attack without defence]
   {\includegraphics[width=0.32\linewidth]{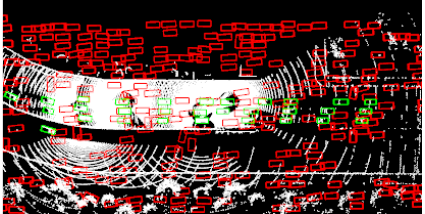} \label{fig:nodefense-BIM}}
    \subfigure[PB attack with ROBOSAC]
    {\includegraphics[width=0.32\linewidth]{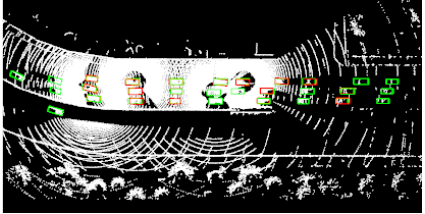} \label{fig:robosac-PB}}
     \subfigure[PGD attack with ROBOSAC]
    {\includegraphics[width=0.32\linewidth]{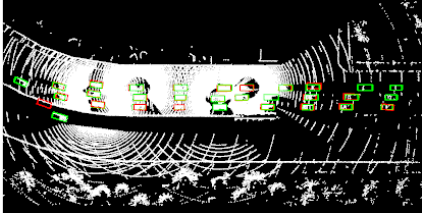} \label{fig:robosac-PGD}}
     \subfigure[BIM attack with ROBOSAC]
    {\includegraphics[width=0.32\linewidth]{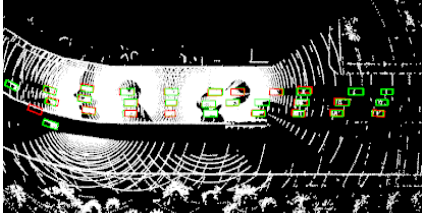} \label{fig:robosac-BIM}}
     \subfigure[PB attack with LUCIA]
    {\includegraphics[width=0.32\linewidth]{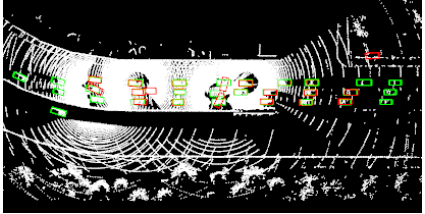} \label{fig:lucia-PB}}
     \subfigure[PGD attack with LUCIA]
    {\includegraphics[width=0.32\linewidth]{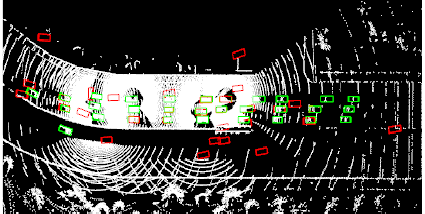} \label{fig:lucia-PGD}}
     \subfigure[BIM attack with LUCIA]
    {\includegraphics[width=0.32\linewidth]{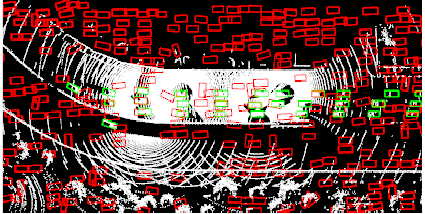} \label{fig:lucia-BIM}}
    \subfigure[PB attack with GLST]
    {\includegraphics[width=0.32\linewidth]{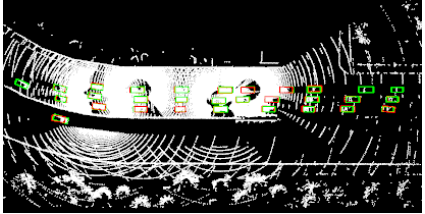} \label{fig:GLST-PB}}
    \subfigure[PGD attack with GLST]
    {\includegraphics[width=0.32\linewidth]{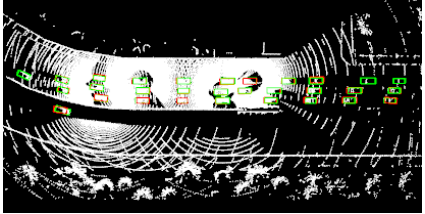} \label{fig:GLST-PGD}}
    \subfigure[BIM attack with GLST]
    {\includegraphics[width=0.32\linewidth]{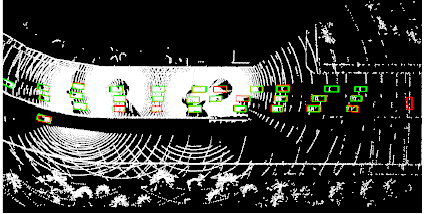} \label{fig:GLST-BIM}}
     }
  \caption{3D detection results of CP systems under PB, PGD, and BIM attacks with different defense methods, including ROBOSAC, LUCIA, and GLST. }
    \label{fig:detection_results}
\end{figure*}

Fig.~\ref{fig:multi_attacker_performance} further evaluates attack effectiveness under the multi-attacker setting. As the number of attackers increases from two to four, the detection performance of Where2comm consistently deteriorates under all three attacks. PGD and BIM remain highly destructive in multi-attacker scenarios, with AP values approaching zero under most IoU thresholds. This indicates that once multiple malicious collaborators inject optimized perturbations into the shared feature space, the collaborative fusion process can be severely disrupted. PB exhibits a more gradual but still clear degradation trend. Specifically, as the number of PB attackers increases from two to four, AP@0.5 decreases from 0.22 to 0.17 and then to 0.09, while AP@0.7 drops from 0.07 to 0.05 and then to 0.02. These results show that PB can continuously weaken the perception capability of Where2comm as the number of attackers increases. Therefore, Fig.~\ref{fig:single_attacker_performance} and Fig.~\ref{fig:multi_attacker_performance} demonstrate that all three attacks are effective against Where2comm, while PGD and BIM achieve stronger raw attack strength and PB still provides substantial attack effectiveness under both single-attacker and multi-attacker settings.

\emph{However, attack effectiveness alone is insufficient to evaluate the practical threat of an adversarial attack. An aggressive attack may significantly reduce AP but also produce many visually abnormal false detections, making it easier to identify and filter.} Therefore, we further analyze attack stealthiness as reported in Table~\ref{tab:fp_iou}. Under the no-attack setting, the false-positive value remains 0.87. PGD produces substantially more false positives, increasing this value to 29.81, 35.00, 45.19, and 42.43 as the number of attackers increases from one to four. BIM is even more aggressive, producing extremely large false-positive values of 321.22, 315.35, 303.77, and 266.62. These results indicate that although PGD and BIM are highly destructive, they introduce a large number of abnormal detections and therefore exhibit weak stealthiness.

\begin{table}[htbp]
  \centering
  \caption{FP$_{\text{IoU0.5}}$ under Different Attacks and Number of Attackers}
  \begin{tabular}{ccccc}
    \toprule
    \multirow{2}{*}{Attack Method} & \multicolumn{4}{c}{Number of Attackers} \\
    \cmidrule{2-5}
    & 1 & 2 & 3 & 4 \\
    \midrule
    No Attack & 0.87 & 0.87 & 0.87 & 0.87 \\
    PGD       & 29.81 & 35 & 45.19 & 42.43 \\
    BIM       & 321.22 & 315.35 & 303.77 & 266.62 \\
    PB        & \textbf{2.65} & \textbf{4.09} & \textbf{4.95} & \textbf{6.04} \\
    \bottomrule
  \end{tabular}
  \label{tab:fp_iou}
\end{table}

\begin{table*}[h]
\centering
\caption{Detection performance of different defense methods under PB attacks with multiple attackers.}
\begin{tabular*}{\linewidth}{@{\extracolsep{\fill}} l *{12}{c} @{}}
\toprule
\multirow{3}{*}{Attack Method}
& \multicolumn{12}{c}{Attacker Number} \\
\cmidrule{2-13}
& \multicolumn{3}{c}{1} & \multicolumn{3}{c}{2} & \multicolumn{3}{c}{3} & \multicolumn{3}{c}{4} \\
\cmidrule{2-4} \cmidrule{5-7} \cmidrule{8-10} \cmidrule{11-13}
& AP0.3 & AP0.5 & AP0.7 & AP0.3 & AP0.5 & AP0.7 & AP0.3 & AP0.5 & AP0.7 & AP0.3 & AP0.5 & AP0.7 \\
\midrule
No Defense
& 0.42 & 0.37 & 0.18
& 0.30 & 0.22 & 0.07
& 0.23 & 0.17 & 0.05
& 0.15 & 0.09 & 0.02 \\
ROBOSAC
& 0.53 & 0.49 & 0.26
& 0.79 & 0.75 & 0.52
& 0.67 & 0.66 & 0.45
& 0.50 & 0.50 & 0.36 \\
LUCIA
& \textbf{0.87} & \textbf{0.85} & \textbf{0.65}
& 0.31 & 0.25 & 0.09
& 0.20 & 0.13 & 0.03
& 0.18 & 0.11 & 0.02 \\
GLST(Ours)
& 0.84 & 0.82 & 0.61
& \textbf{0.82} & \textbf{0.79} & \textbf{0.57}
& \textbf{0.70} & \textbf{0.67} & \textbf{0.46}
& \textbf{0.69} & \textbf{0.67} & \textbf{0.50} \\
\bottomrule
\end{tabular*}
\label{tab:pb_multi_attackers}
\end{table*}

\begin{figure*}[t]
    \centering
    {
    \subfigure[No Attack]
     {\includegraphics[width=0.24\linewidth]{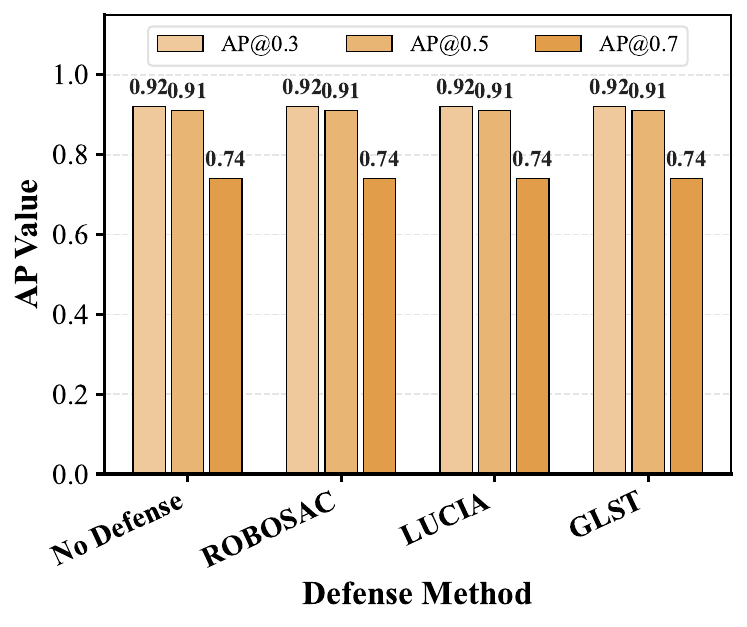} \label{fig:defense_single_No_Attack}}
     \subfigure[PGD]
   {\includegraphics[width=0.24\linewidth]{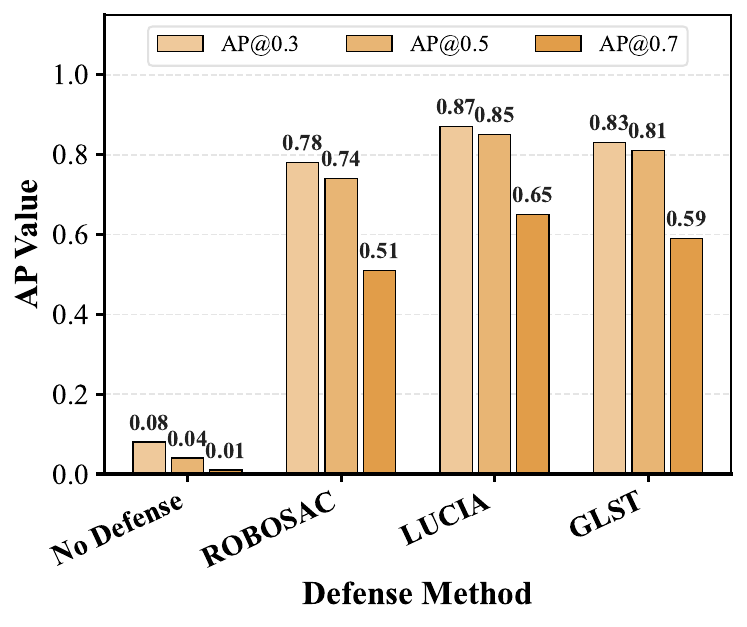}\label{fig:defense_single_PGD}}
    \subfigure[BIM]
   {\includegraphics[width=0.24\linewidth]{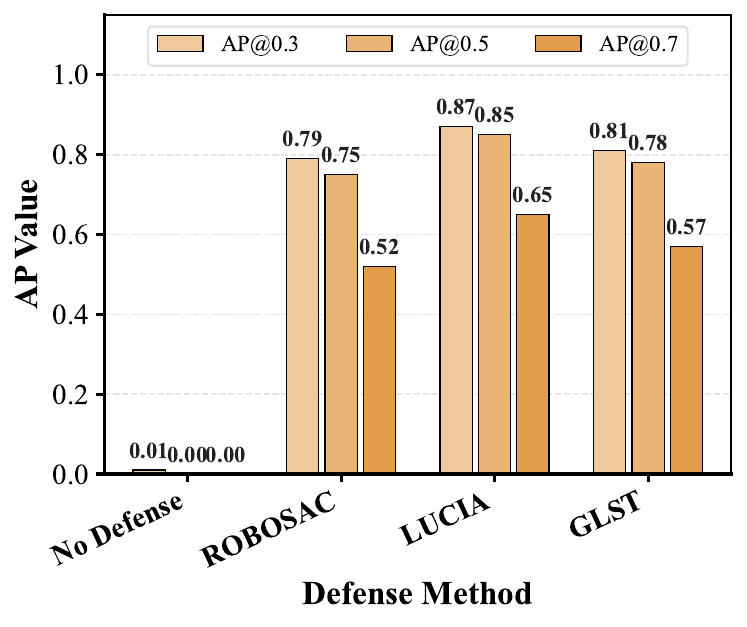} \label{fig:defense_single_BIM}}
    \subfigure[PB]
    {\includegraphics[width=0.24\linewidth]{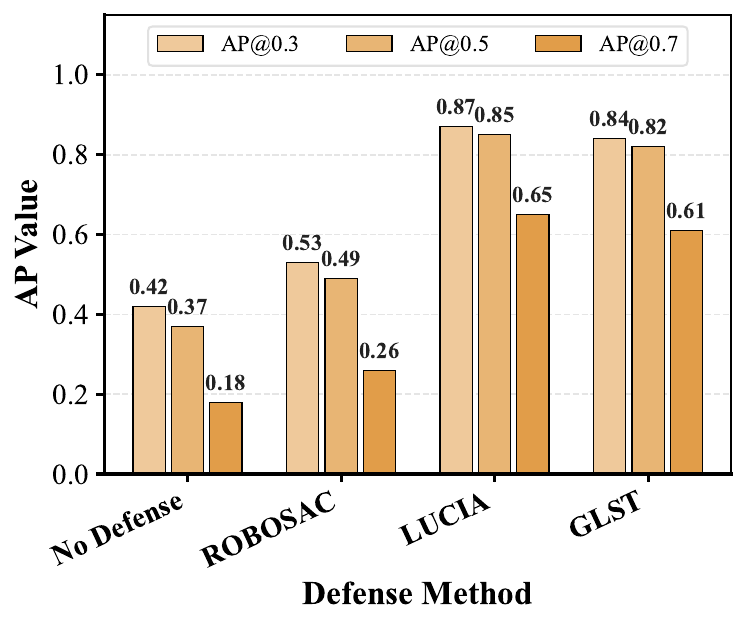} \label{fig:defense_single_PB}}
     }
\caption{Defense performance of ROBOSAC, LUCIA, and GLST against PGD, BIM, and PB attacks under the single-attacker setting.}
\label{fig:defense_single_attacker}
\end{figure*}

In contrast, PB produces far fewer false positives than PGD and BIM. Its false-positive value increases only from 2.65 to 6.04 as the number of attackers increases from one to four, remaining much closer to the no-attack case than the other two attacks. This suggests that PB can preserve relatively benign-looking detection behavior while still causing substantial AP degradation. Combining the AP results in Fig.~4 and Fig.~5 with the stealthiness results in Table~I, we can observe a clear distinction among the three attacks. PGD and BIM mainly emphasize destructive capability but suffer from poor stealthiness due to excessive false positives. PB, on the other hand, achieves a better balance between attack effectiveness and stealthiness: it significantly degrades cooperative 3D object detection while producing much fewer abnormal false detections. \emph{This makes PB a more practical and challenging threat model for evaluating the robustness of secure CP defenses}.

\subsection{Defense Performance Analysis Of GLST}

To evaluate the defense capability of the proposed GLST framework, we compare it with two representative CP defense methods, ROBOSAC and LUCIA, under both single-attacker and multi-attacker scenarios. We first provide a qualitative comparison in Fig.~\ref{fig:detection_results}. Without defense, adversarial attacks introduce severe perception errors, including false positives, missed detections, and shifted bounding boxes. After applying GLST, most attack-induced false detections are effectively suppressed, and the predicted bounding boxes become much more consistent with the ground-truth annotations. This visual comparison indicates that GLST can reduce the influence of malicious intermediate features and recover reliable cooperative perception results under different attack types.

Fig.~\ref{fig:defense_single_attacker} reports the quantitative defense performance under the single-attacker setting. In this relatively simple scenario, both LUCIA and GLST achieve strong recovery performance against PGD, BIM, and PB attacks. Under the PB attack, GLST improves AP@0.5 from 0.37 without defense to 0.82, showing that the proposed trust reweighting mechanism can effectively suppress stealthy malicious features. Although LUCIA achieves slightly higher AP values in some single-attacker cases, GLST remains highly competitive with the strongest baseline. This result suggests that the proposed multi-level trust modeling does not sacrifice single-attacker defense capability while preparing the system for more complex adversarial settings.

\begin{table*}[t]
\centering
\caption{Detection performance of different defense methods under PGD attacks with multiple attackers.}
\begin{tabular*}{\linewidth}{@{\extracolsep{\fill}} l *{12}{c} @{}}
\toprule
\multirow{3}{*}{Attack Method}
& \multicolumn{12}{c}{Attacker Number} \\
\cmidrule{2-13}
& \multicolumn{3}{c}{1} & \multicolumn{3}{c}{2} & \multicolumn{3}{c}{3} & \multicolumn{3}{c}{4} \\
\cmidrule{2-4} \cmidrule{5-7} \cmidrule{8-10} \cmidrule{11-13}
& AP0.3 & AP0.5 & AP0.7 & AP0.3 & AP0.5 & AP0.7 & AP0.3 & AP0.5 & AP0.7 & AP0.3 & AP0.5 & AP0.7 \\
\midrule
No Defense
& 0.08 & 0.04 & 0.01
& 0.02 & 0.01 & 0.00
& 0.01 & 0.00 & 0.00
& 0.00 & 0.00 & 0.00 \\
ROBOSAC
& 0.78 & 0.74 & 0.51
& 0.79 & 0.75 & 0.52
& 0.76 & 0.70 & 0.47
& \textbf{0.73} & \textbf{0.68} & \textbf{0.49} \\
LUCIA
& \textbf{0.87} & \textbf{0.85} & \textbf{0.65}
& 0.08 & 0.05 & 0.02
& 0.03 & 0.01 & 0.00
& 0.01 & 0.00 & 0.00 \\
GLST(Ours)
& 0.83 & 0.81 & 0.59
& \textbf{0.81} & \textbf{0.78} & \textbf{0.55}
& \textbf{0.82} & \textbf{0.78} & \textbf{0.55}
& 0.72 & 0.68 & 0.49 \\
\bottomrule
\end{tabular*}
\label{tab:pgd_multi_attackers}
\end{table*}

The core advantage of GLST becomes more evident in multi-attacker scenarios. Table~\ref{tab:pb_multi_attackers} reports the defense performance under PB attacks with different numbers of malicious agents. When no defense is applied, the perception performance continuously decreases as the number of PB attackers increases, with AP@0.5 dropping from 0.37 under one attacker to 0.09 under four attackers. LUCIA performs well in the single-attacker setting, achieving 0.85 AP@0.5. However, its performance rapidly collapses when multiple attackers are present. Specifically, AP@0.5 decreases to 0.25, 0.13, and 0.11 when the number of attackers increases to two, three, and four, respectively. This confirms that pairwise distance-based trust estimation is vulnerable to the pseudo-consensus formed by multiple malicious agents.

In contrast, GLST maintains stable robustness under multi-attacker PB attacks. When the number of attackers increases from two to four, GLST achieves 0.79, 0.67, and 0.67 AP@0.5, respectively. Even in the four-attacker scenario, GLST still maintains 0.69 AP@0.3, 0.67 AP@0.5, and 0.50 AP@0.7, significantly outperforming LUCIA and also exceeding ROBOSAC in most metrics. These results demonstrate that the proposed method can effectively mitigate the consensus bias caused by coordinated attackers.

\begin{figure}[t]
    \centering
    {
    \subfigure[LUCIA trust score]
     {\includegraphics[width=0.47\linewidth]{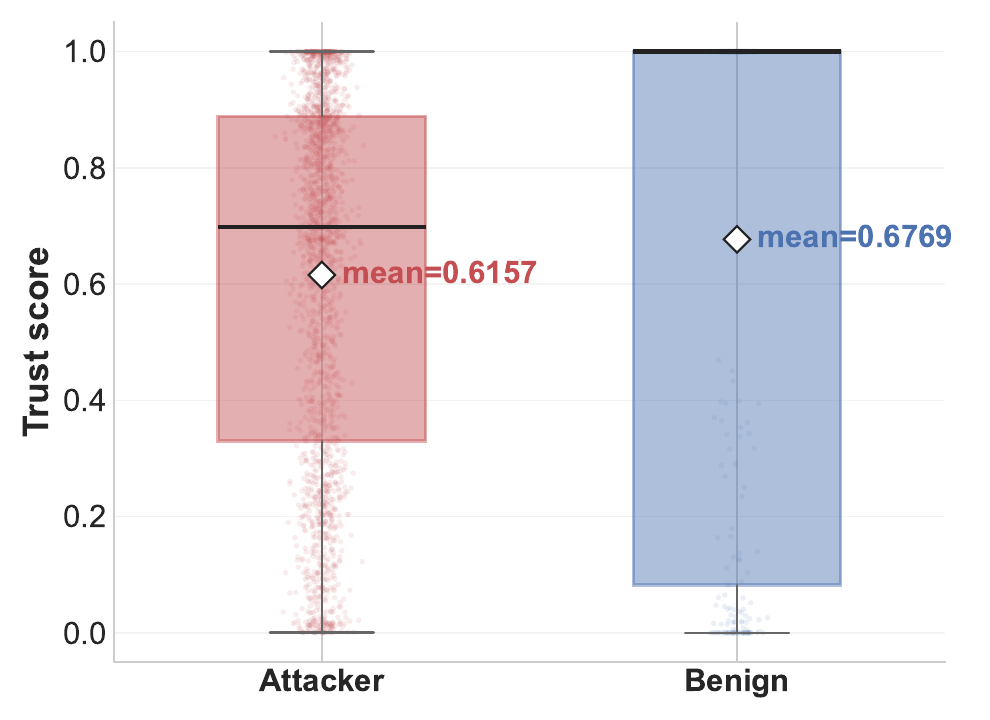} \label{fig:LUCIA_trust}}
     \subfigure[GLST trust score]
   {\includegraphics[width=0.47\linewidth]{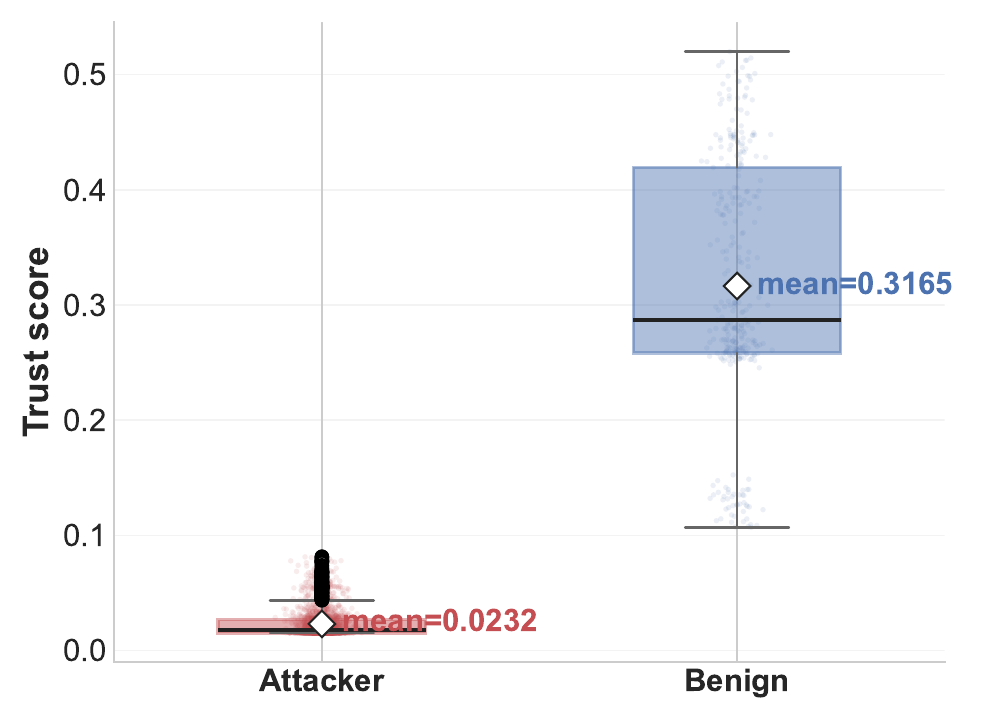}\label{fig:GLST_trust}}
     }
    \caption{Box plot of trust score distribution for defense algorithms.}
    \label{fig:trust_scores}
\end{figure}

To further examine whether GLST is specific to PB attacks, we also evaluate its robustness under multi-attacker PGD attacks, as shown in Table~\ref{tab:pgd_multi_attackers}. Without defense, the detector almost completely fails under PGD, with AP values approaching zero as the number of attackers increases. LUCIA again shows strong performance in the single-attacker setting but degrades sharply under multiple attackers, indicating that its single-signal trust estimation is insufficient when malicious agents reinforce each other in the feature space. ROBOSAC maintains relatively stable performance under PGD attacks, but its sampling-based strategy introduces additional computational cost. GLST achieves comparable or better performance in most multi-attacker PGD settings. For example, under two and three PGD attackers, GLST maintains 0.78 AP@0.5, and under four attackers it still achieves 0.68 AP@0.5. These results indicate that GLST is not tailored only to PB attacks; it also generalizes well to strong gradient-based feature attacks.

Finally, Fig.~\ref{fig:trust_scores} provides an explanation for the superior robustness of GLST from the perspective of trust-score distributions. In the two-PB-attacker scenario, LUCIA assigns highly overlapping trust scores to benign and malicious agents, with mean scores of 0.6769 and 0.6157, respectively. Such a small separation makes it difficult to distinguish attackers from benign collaborators, especially when malicious agents form a pseudo-consensus. In contrast, GLST assigns a much lower mean trust score to attackers than to benign agents, with mean scores of 0.0232 and 0.3165, respectively. This clear separation shows that GLST provides stronger discriminability between benign and malicious collaborators. Therefore, the performance gain of GLST comes not merely from feature reweighting, but from more reliable trust estimation enabled by the joint modeling of global, local, and structural consistency.

Overall, the experimental results demonstrate that GLST achieves competitive defense performance under single-attacker scenarios and substantially stronger robustness under multi-attacker scenarios. Compared with defenses relying on a single consistency signal, GLST is more resistant to stealthy feature injection and multi-attacker pseudo-consensus. These findings validate the necessity of multi-level trust modeling for secure confidence-driven collaborative perception.

\subsection{Ablation Study}

\begin{figure}[t]
    \centering
    {
    \subfigure[Single-attacker]
     {\includegraphics[width=0.47\linewidth]{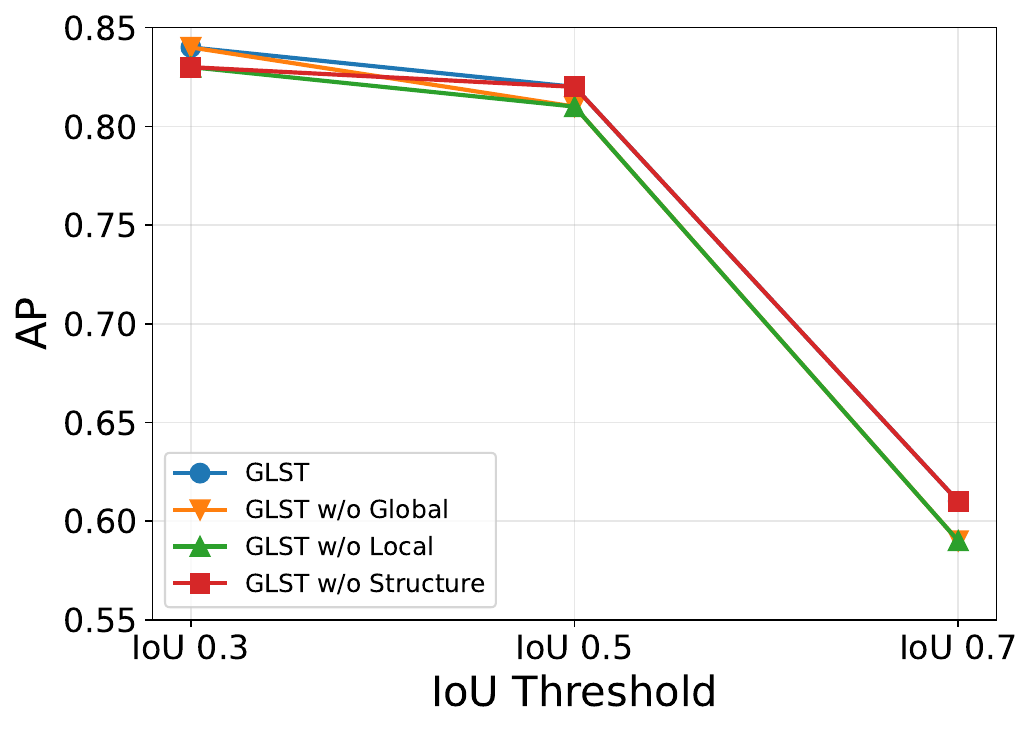} \label{fig:ablation_1attacker}}
     \subfigure[Dual-attacker]
     {\includegraphics[width=0.47\linewidth]{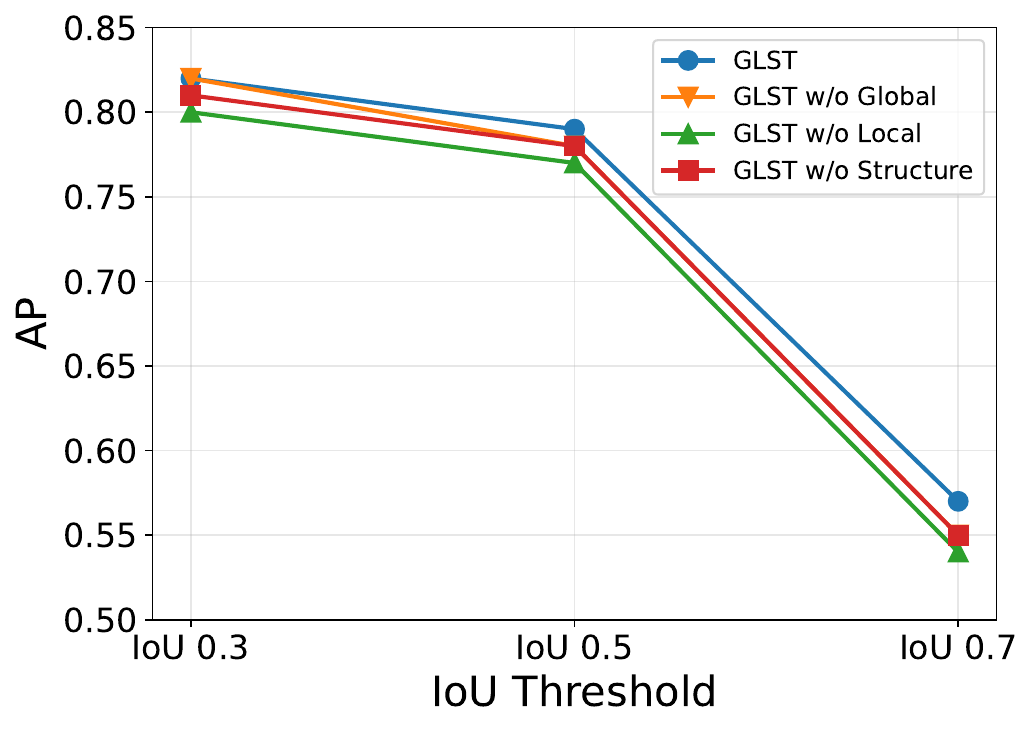}\label{fig:ablation_2attacker}}
     }
  \caption{Ablation study of the three main components in the proposed GLST.}
  \label{fig:ablation}
\end{figure}

To further analyze the contribution of each trust component in GLST, we conduct ablation studies under PB attacks are shown in Fig.~\ref{fig:ablation}. Three variants are considered: GLST w/o Global, which removes the global consistency branch; GLST w/o Local, which removes the multi-scale local residual branch; and GLST w/o Structure, which removes the structural consistency branch.

As shown in Fig.~\ref{fig:ablation_1attacker}, under the single-attacker setting, all variants achieve relatively close performance, indicating that each remaining pair of trust cues can still provide useful defense information when the attack setting is simple. Nevertheless, the full GLST consistently achieves the best overall performance across all IoU thresholds. This result shows that global consistency, local residual consistency, and structural consistency are complementary rather than redundant. The advantage of the full GLST becomes more evident in the dual-attacker setting, as shown in Fig.~\ref{fig:ablation_2attacker}. When multiple attackers are present, removing any single branch leads to a clear performance drop, especially under stricter IoU thresholds. Among the three variants, GLST w/o Local suffers the largest degradation at AP@0.7, suggesting that the multi-scale local residual branch plays a key role in detecting fine-grained adversarial perturbations in perception-critical regions. This is consistent with the design motivation of GLST, since stealthy attacks such as PB tend to introduce localized feature manipulations while preserving benign-like global statistics.

\section{Conclusion}

In this paper, we investigated the security of confidence-driven V2X collaborative perception against stealthy multi-attacker feature-injection attacks. Using Where2comm as a representative framework, we showed that PB attacks can exploit spatial confidence-based communication by injecting benign-like adversarial features into perception-critical regions, thereby significantly degrading 3D object detection while maintaining strong stealthiness. To address this threat, we proposed GLST, a lightweight global-local-structural trust framework that jointly evaluates collaborator reliability from global feature consistency, multi-scale local residual consistency, and structural consistency with ego-side semantic topology. The resulting trust scores are incorporated into the feature fusion process to suppress unreliable collaborators. Experiments demonstrate that GLST achieves competitive defense performance under single-attacker scenarios and substantially stronger robustness under multi-attacker PB and PGD attacks. Ablation results further confirm that the three trust components provide complementary information and jointly contribute to reliable defense against stealthy and coordinated malicious collaborators.

\bibliographystyle{IEEEtran}
\bibliography{ieeetran}

\end{document}